\documentclass[11pt]{article}
\usepackage{jheppub}
\usepackage{amsmath,amssymb,amsfonts,graphicx}

\newcommand{\be}{\begin{equation}}
\newcommand{\ee}{\end{equation}}
\newcommand{\bea}{\begin{eqnarray}}
\newcommand{\eea}{\end{eqnarray}}
\newcommand{\beas}{\begin{eqnarray*}}
\newcommand{\eeas}{\end{eqnarray*}}
\newcommand{\ba}{\begin{array}}
\newcommand{\ea}{\end{array}}

\renewcommand*\d[2][]{%
	\mathrm{d}%
	\ifx\relax#1\relax\else
	\rule{-0.02em}{1.5ex}^{#1}\rule{0.08em}{0ex}\!
	\fi
	#2\,
}

\title{Cosmology from the vacuum}

\author[1]{Stefano Antonini,}
\author[2]{Petar Simidzija,}
\author[1,3]{Brian Swingle,}
\author[2]{Mark Van Raamsdonk}

\affiliation[1]{Maryland Center for Fundamental Physics, University of Maryland, College Park, MD 20742, USA}
\affiliation[2]{Department of Physics and Astronomy, University of British Columbia,\\
6224 Agricultural Road, Vancouver, B.C.\ V6T 1Z1, Canada.}
 \affiliation[3]{Brandeis University, Waltham, MA 02453, USA}

\emailAdd{santonin@umd.edu}
\emailAdd{psimidzija@phas.ubc.ca}
\emailAdd{bswingle@brandeis.edu}
\emailAdd{mav@phas.ubc.ca}

\abstract{We argue that standard tools of holography can be used to describe fully non-perturbative microscopic models of cosmology in which a period of accelerated expansion may result from the positive potential energy of time-dependent scalar fields evolving towards a region with negative potential. In these models, the fundamental cosmological constant is negative, and the universe eventually recollapses in a time-reversal symmetric way. The microscopic description naturally selects a special state for the cosmology. In this framework, physics in the cosmological spacetime is dual to the vacuum physics in a static planar asymptotically AdS Lorentzian wormhole spacetime, in the sense that the background spacetimes and observables are related by analytic continuation. The dual spacetime is weakly curved everywhere, so any cosmological observables can be computed in the dual picture via effective field theory without detailed knowledge of the UV completion or the physics near the big bang. In particular, while inflation may explain the origin of perturbations in the cosmology picture, the perturbations can be deduced from the dual picture without any knowledge of the inflationary potential.}

\keywords{}

\begin{document}

\maketitle
\newpage
\parskip=10pt

\section{Introduction}

Holographic models from string theory \cite{Maldacena:1997re,Gubser1998,Witten1998,Aharony1999} have been extremely successful in describing a great variety of quantum gravitational phenomena, including the physics of high-energy scattering processes where gravitational effects are important, the formation, dynamics, and evaporation of black holes, and even the emergence of spacetime and gravitation from intrinsically quantum mechanical phenomena. 
However, it remains a significant open challenge to find a quantum gravitational description of the physics in cosmological spacetimes like our own universe.

\subsubsection*{Cosmology with $\Lambda < 0$}

A possible misconception is that the obstruction to describing realistic cosmological physics using holography is that holographic models naturally describe spacetimes with a negative cosmological constant while our own universe appears to have a positive cosmological constant. However, the observational evidence \cite{perlmutter1999measurements, riess1998observational} tells us not that there is a positive cosmological constant, but that the expansion of the universe is currently accelerating, i.e. that $\ddot{a} > 0$ where $a$ is the scale factor. Einstein's equations then imply that $3p + \rho < 0$ where $p$ and $\rho$ are the pressure and energy density. This could be explained {\it either} by a positive cosmological constant, {\it or} by the potential energy associated with time-dependent scalar fields that are currently at a positive value of their potential \cite{Peebles:1987ek,Ratra:1987rm,Caldwell:1997ii}.\footnote{In the latter case, there are also constraints on how quickly the scalar can be changing. There may be additional possibilities based on strongly coupled quantum fields or modified gravity.} While the $\Lambda$CDM model with $\Lambda>0$ provides a good fit to cosmological data, the possibility of $\Lambda<0$ with time-dependent scalars is not ruled out (see \cite{Dutta:2018vmq,Visinelli:2019qqu,sen2021cosmological} for recent discussions). Time-dependent scalars are also natural to consider from an effective field theory point of view since they preserve the same symmetries as an FRW spacetime. In this case, there is no particular reason why the scalars can't be evolving towards a minimum or extremum of the potential with a value (which we can define to be the fundamental cosmological constant in the effective theory) that is zero or negative.

In the absence of current observations that can tell us more definitively whether the fundamental cosmological constant is positive, negative, or zero, we might ask which possibility is most natural from a theoretical perspective. The current situation in string theory is that we have (via holography) a wealth of deeply understood models that describe quantum gravitational physics with a negative cosmological constant\footnote{More specifically, we have models where the magnitude of this is similar to the observed magnitude and the size of the extra dimensions are acceptably small. Furthermore, these models typically have scalars whose potential has derivatives of scale similar to that of the cosmological constant - this is equivalent to having one or more scalar operators of low dimension in the dual CFT. Thus, there are likely to be nearby points on the potential with small positive values and a small enough slope so that the scalars change acceptably slowly.} and no fully microscopic models of four-dimensional quantum gravity with zero or positive cosmological constant.\footnote{See \cite{Banks:2001px,Strominger:2001pn,Alishahiha:2004md,Gorbenko:2018oov,Coleman:2021nor,Freivogel2005,McFadden:2009fg,Banerjee:2018qey,Susskind:2021dfc} for various approaches to describing $\Lambda > 0$ cosmology.} Even at the level of effective field theory, it is still controversial whether metastable string compactifications with $\Lambda > 0$ exist (see e.g. \cite{Obied:2018sgi,Danielsson:2018ztv} for recent discussions). So at present, the possibility that seems most likely to admit a fully non-perturbative theoretical description based on our current understanding is $\Lambda < 0$.\footnote{Of course, it could just be a historical accident that we understood quantum gravity with $\Lambda < 0$ at least 25 years before $\Lambda > 0$.}

Given that we already have a good theoretical understanding of other types of spacetimes with $\Lambda < 0$ through holography, 
a natural next step toward understanding cosmology from a quantum gravity point of view is to try to understand $\Lambda < 0$ cosmologies microscopically, especially since it seems possible that our universe could have $\Lambda < 0$.\footnote{This was also emphasized recently in \cite{Visinelli:2019qqu}. It is certainly still interesting to understand whether string theory can produce $\Lambda > 0$ models and whether these can be given a non-perturbative description, but it seems to us that comparatively little effort has been expended exploring what may be a simpler possibility.} 

\subsubsection*{Why it is hard to describe $\Lambda < 0$ cosmological models holographically}

Even if we presume $\Lambda < 0$, there is still a significant obstacle in using holography to describe models of cosmology: cosmological models assume a homogeneous and isotropic universe filled with matter and radiation, while usual holographic models describe spacetimes that are {\it asymptotically empty}. While holography has no trouble describing spacetimes with black holes, gravitational waves, radiation, etc... these interesting features are always in the interior of a spacetime that asymptotically looks just like pure AdS.\footnote{We can also have other types of rigid asymptotic behaviors in models where the dual field theory is not conformal in the UV.} This rigid asymptotic behaviour is related to the fact that any finite energy state of the dual field theory has the same UV physics (e.g. short distance correlators).

A complementary way to view the difficulty is that cosmological spacetimes with $\Lambda < 0$ are typically big-bang / big-crunch geometries with no asymptotically AdS regions that could be associated with a dual field theory.

\subsubsection*{Asymptocially AdS regions in complex time}

\begin{figure}
  \centering
  \includegraphics[scale=0.25]{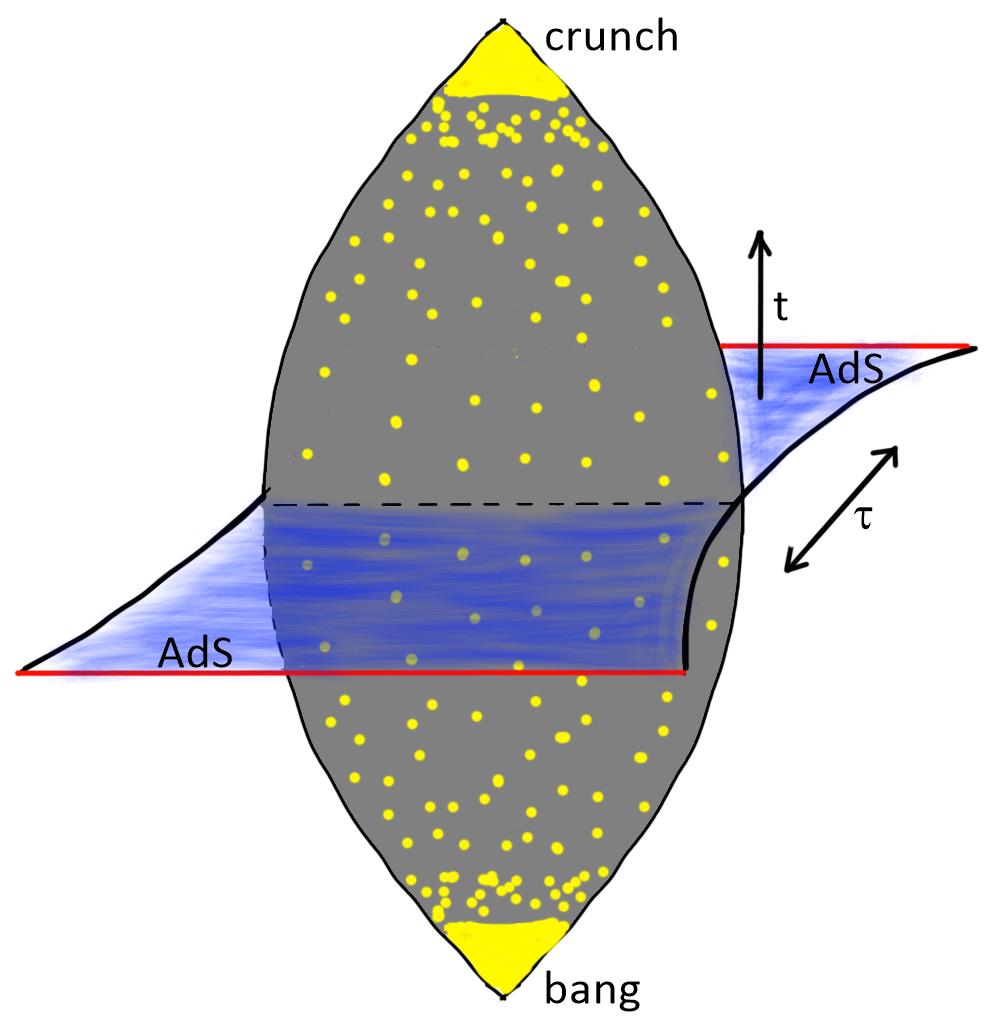}
  \caption{Asymptotically AdS regions in the analytic continuation of a time-symmetric $\Lambda < 0$ cosmology. The asymptotically AdS Euclidean gravity theory can be given a microscopic holographic description, and this can be used to define a special state for the cosmology.}
\label{fig:analytic}
\end{figure}

A hint for how to make progress is that these $\Lambda < 0$ cosmological spacetimes often do have asymptotically AdS regions if we consider complexifying the coordinates. In an expanding universe with $\Lambda < 0$, the negative cosmological constant eventually comes to dominate the evolution, and this causes the scale factor to decelerate and eventually turn around. In cases where any scalar fields also have zero time derivative at this point, the full solution for the background geometry is time-reversal symmetric about this turn-around point. In such time-symmetric geometries, we can analytically continue the time direction to obtain a real Euclidean geometry.\footnote{The scale factor is typically analytic around the time-reversal symmetric point since it is a solution of the Friedmann equation, and the analytically continued Friedmann equation is sensible for typical equations of state.} This analytically continued geometry has a pair of asymptotically AdS regions at Euclidean time $ \tau = \pm \infty$ (see Figure \ref{fig:analytic}) provided that the scalar fields asymptote to some value with a negative potential. In the special case of a flat cosmology (that we will focus on), we can perform a further analytic continuation of one of the spatial directions to obtain a {\it static} Lorentzian geometry with a pair of asymptotically AdS regions.

So far, this is a mathematical curiosity: starting from a  time-reversal symmetric $\Lambda < 0$ flat cosmological spacetime, we can associate via analytic continuation a Euclidean geometry and a static horizon-free Lorentzian geometry each with a pair of asymptotically AdS regions.

However, the observation suggests a path forward to find a microscopic description of the $\Lambda < 0$ cosmology:
\begin{enumerate}
\item
Find a microscopic description for the physics of the associated spacetimes with asymptotically AdS regions using holography.
\item
Use this microscopic description of the Euclidean theory to define a special state for the Lorentzian cosmology and to give a holographic description of the physics of this state.
\end{enumerate}
Before turning to the holographic descriptions, let us further motivate the second point here.

\subsubsection*{A special state for cosmology}

In quantum field theory, if we have a time-reversal symmetric background Lorentzian spacetime related by analytic continuation to a reflection-symmetric Euclidean spacetime, the quantum field theory on this Euclidean spacetime\footnote{The Euclidean field theory may require additional boundary conditions for fields to complete its definition, e.g. if the analytically continued spacetime has asymptotically AdS regions. In this case, we have a family of states, each associated with a choice of boundary conditions.} defines a special state of the Lorentzian theory, obtained by slicing the path integral defining the Euclidean theory (see Appendix A for a review). Observables for this state in the Lorentzian theory are related by analytic continuation to observables in the Euclidean theory. For example, for a field theory in Minkowski space, the state defined by the associated Euclidean field theory is the vacuum state.

In our case, the spacetimes are dynamical, but we will argue that the Euclidean version of the gravitational theory describing the Lorentzian cosmology can still be used to define a preferred quantum state for the cosmology. This is similar to the idea of Hartle and Hawking \cite{Hartle:1983ai}, who argued that a natural state for cosmology is provided by a Euclidean gravitational path integral with no boundary in the Euclidean past. 

In our case, the Euclidean spacetimes we consider do have a boundary in the Euclidean past, namely an asymptotically AdS boundary. This is important for us as it will permit us to define the gravitational path integral non-perturbatively using holography. 

\subsubsection*{Holographic description of spacetimes with two asymptotically AdS regions}

We now return to the question of how to holographically describe the physics of these spacetimes. For our discussion, we focus on the most phenomenologically relevant case of four-dimensional flat cosmology; we comment on the generalization to other cases in the discussion.

We have seen that for a large class of $\Lambda < 0$ flat cosmological spacetimes, analytic continuation gives a Euclidean geometry of the form
\be
ds^2 = d \tau^2 + a^2(\tau)(dx^i dx^i)
\ee
with asymptotically AdS regions for $\tau \to \pm \infty$. A further analytic continuation gives a Lorentzian static spacetime
\be
ds^2 = d \tau^2 + a^2(\tau)(dx_\mu dx^\mu) \; .
\ee
Such geometries are sometimes described as asymptotically AdS planar {\it wormholes}, since they have two asymptotically AdS regions connected by a ``throat'' with a planar cross section. The two asymptotically AdS regions here suggest a dual holographic description that involves a pair of three-dimensional CFTs,\footnote{The AdS/CFT description of Euclidean wormholes was first discussed in \cite{Maldacena:2004rf}, where the possible connection to cosmological physics via analytic continuation was also pointed out. More recent discussions of the holographic description of the types of wormholes we are discussing here appears in \cite{Freivogel:2019lej, Betzios:2019rds}.} where the coordinates labeled by $x$ correspond to the directions on which the field theory lives, and the $\tau$ direction is the emergent radial direction. 

\begin{figure}
  \centering
  \includegraphics[scale=0.35]{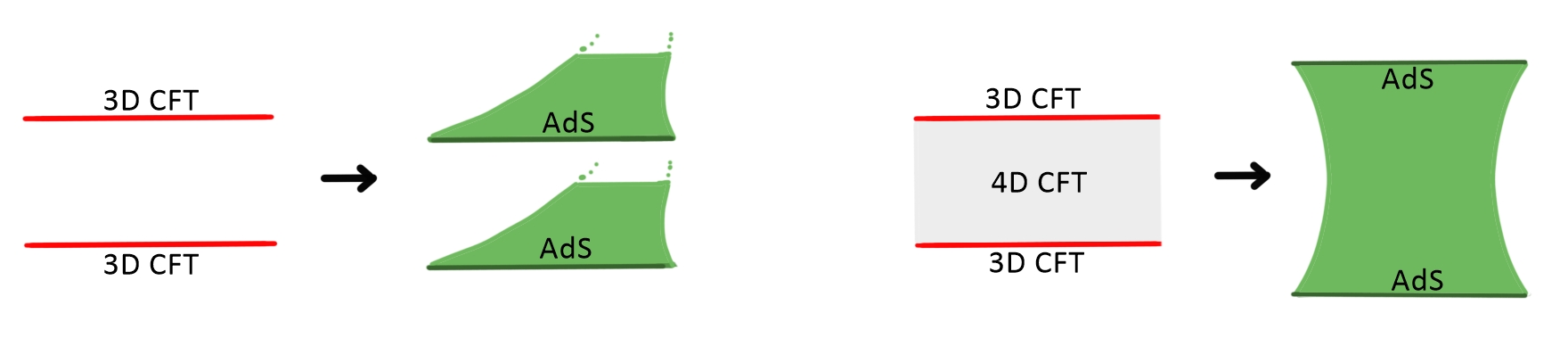}
  \caption{Left: two non-interacting 3D CFTs are dual to a pair of disconnected AdS spacetimes. Right: two 3D CFTs interacting via a non-holographic 4D CFT, proposed to be dual (in some cases) to a Lorentzian wormhole with two asymptotically AdS boundaries.}
\label{fig:coupling}
\end{figure}


For the Lorentzian wormhole, since it is possible to  go (causally) from one asymptotically AdS region to the other through the geometry, the two CFTs in the holographic description must be interacting somehow.\footnote{For Euclidean wormhole solutions in general, it has been suggested that the dual description may be in terms of some type of ensemble of CFTs, since the connected geometry suggests correlations between the two sides, but such correlations will be absent for a single pair of non-interacting CFTs \cite{Maldacena:2004rf, Saad:2019lba, Marolf:2020xie, Maloney:2020nni, Afkhami-Jeddi:2020ezh, Cotler:2020ugk}. In the special case where the wormhole cross-section has a translation-invariant direction so that we can analytically continue to a static Lorentzian wormhole, we can make the stronger statement that the CFTs must be interacting \cite{VanRaamsdonk:2020tlr}. As discussed in \cite{VanRaamsdonk:2020tlr}, when this interaction is mediated via auxiliary degrees of freedom, integrating these degrees of freedom out in the Euclidean picture does give rise to something like an ensemble of CFTs, where the ensemble is over possible sources for local operators.} The physical separation between the boundaries in the wormhole geometry indicates that the interactions between the two CFTs are not instantaneous and do not induce arbitrarily strong correlations \cite{Betzios:2019rds,VanRaamsdonk:2020tlr}. These features suggest that the interaction is not a local field theory interaction; in \cite{VanRaamsdonk:2020tlr}, we argued that the interaction likely involves some auxiliary degrees of freedom and that a natural possibility is for these extra degrees of freedom to take the form of a four-dimensional quantum field theory on $\mathbb{R}^3$ times an interval, to which the two CFTs are coupled at the edges of the interval (see Figure \ref{fig:coupling}).\footnote{More generally, the two 3D CFTs could be boundaries or defects in a higher-dimensional quantum field theory. This seems to be the most general possibility consistent with Lorentz invariance, local quantum field theory, and the absence of instantaneous interactions and diverging correlators between the two 3D theories.} The proposed dual for the asymptotically AdS wormhole geometry is the vacuum state of this 3D-4D-3D theory.

To preserve the four-dimensional character of the dual gravitational theory, the four-dimensional field theory should have many fewer local degrees of freedom than the 3D CFTs so that it represents a small perturbation to the physics in the UV. Nevertheless, via a long RG flow, this perturbation leads to a large amount of entanglement between the 3D CFTs in the IR, giving rise to the connected dual geometry \cite{VanRaamsdonk:2020tlr}. We review some additional motivations and evidence for this construction in section 2 below.

\subsubsection*{Holographic description of the cosmological physics}

Assuming this dual description of the Lorentzian wormhole spacetime, we can now try to understand the dual of the cosmological physics.

The proposed dual for the Euclidean wormhole is the Euclidean version of the same field theory: a pair of Euclidean 3D CFTs on $\mathbb{R}^3$ coupled via a Euclidean 4D CFT on $\mathbb{R}^3$ times an interval.\footnote{More precisely, the wormhole is the dominant saddle point geometry in a gravitational path integral dual to this field theory path integral.}. 
This Euclidean theory can be used to construct the Lorentzian vacuum state dual to the Lorentzian wormhole by slicing the path integral perpendicular to the direction that is to be analytically continued (one of the $\mathbb{R}^3$ directions), as shown in Figure \ref{fig:slices}.\footnote{See Appendix \ref{app:slicing} for a review of how quantum states can be defined by slicing Euclidean path integrals.}

\begin{figure}
  \centering
  \includegraphics[scale=0.32]{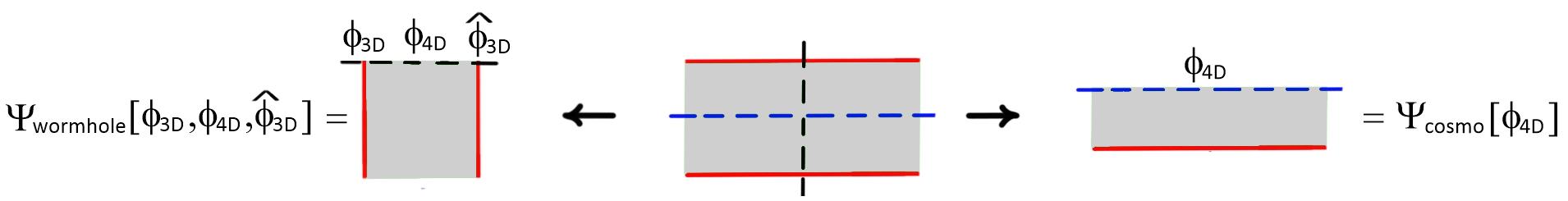}
  \caption{Two different slicings of the same microscopic Euclidean path integral give states dual to the Lorentzian wormhole and the cosmology. The pictures on the left and right indicated a Euclidean path integral over field configurations with the specified boundary conditions at the top surface. See also Figure 6 for the gravity interpretation of these states.}
\label{fig:slices}
\end{figure}

To define a state associated with the cosmological spacetime, we slice the same path integral in a different way, perpendicular to the direction between the two 3D CFTs, along the surface fixed by the reflection symmetry that exchanges the two 3D CFTs. We can understand this via the right-hand figure of figure 2: the sliced CFT path integral corresponds to the sliced gravitational path integral that constructs a state for the cosmology at the time-symmetric point.

This second slicing defines a state of the 4D CFT on spatial $\mathbb{R}^3$. The 3D CFT appears only in the Euclidean path integral used to construct the state, not in the physical degrees of freedom. Even though the 4D CFT is not conventionally holographic, the presence of the holographic 3D CFT in the Euclidean past of the path integral produces a state that is sufficiently complex to encode the semiclassical physics of the cosmological spacetime.

The physics here is similar to how black hole interiors can be encoded in auxiliary radiation systems \cite{Maldacena2013,VanRaamsdonk:2013sza,Penington:2019npb,Almheiri2019b}, or of how bubbles of spacetime associated with a holographic CFT can be encoded in the state of a different CFT \cite{Simidzija:2020ukv} or a collection of non-interacting BCFTs \cite{VanRaamsdonk2018}.

\subsubsection*{Extracting cosmological observables}

The 4D CFT state that encodes the cosmological spacetime is a high-energy state that will look almost completely thermal. Thus, extracting the cosmological observables from this state would appear to be extremely difficult, similar to extracting local behind-the-horizon physics from a CFT state describing a single-sided black hole.\footnote{The entanglement wedge of a large enough subsystem of the 4D CFT will include part of the cosmological spacetime, so entanglement entropy for such large subsystems does provide one direct probe of the cosmology.} Indeed, in the special case where the 4D CFT is holographic, the cosmological physics can be understood as living on an end-of-the-world (ETW) brane behind a black hole horizon.\footnote{This was the case in the first examples of models of this type \cite{Cooper2018, Antonini2019}, where the approach was dubbed ``Black hole microstate cosmology''. See section 2.1 for a review} 

\begin{figure}
  \centering
  \includegraphics[scale=0.32]{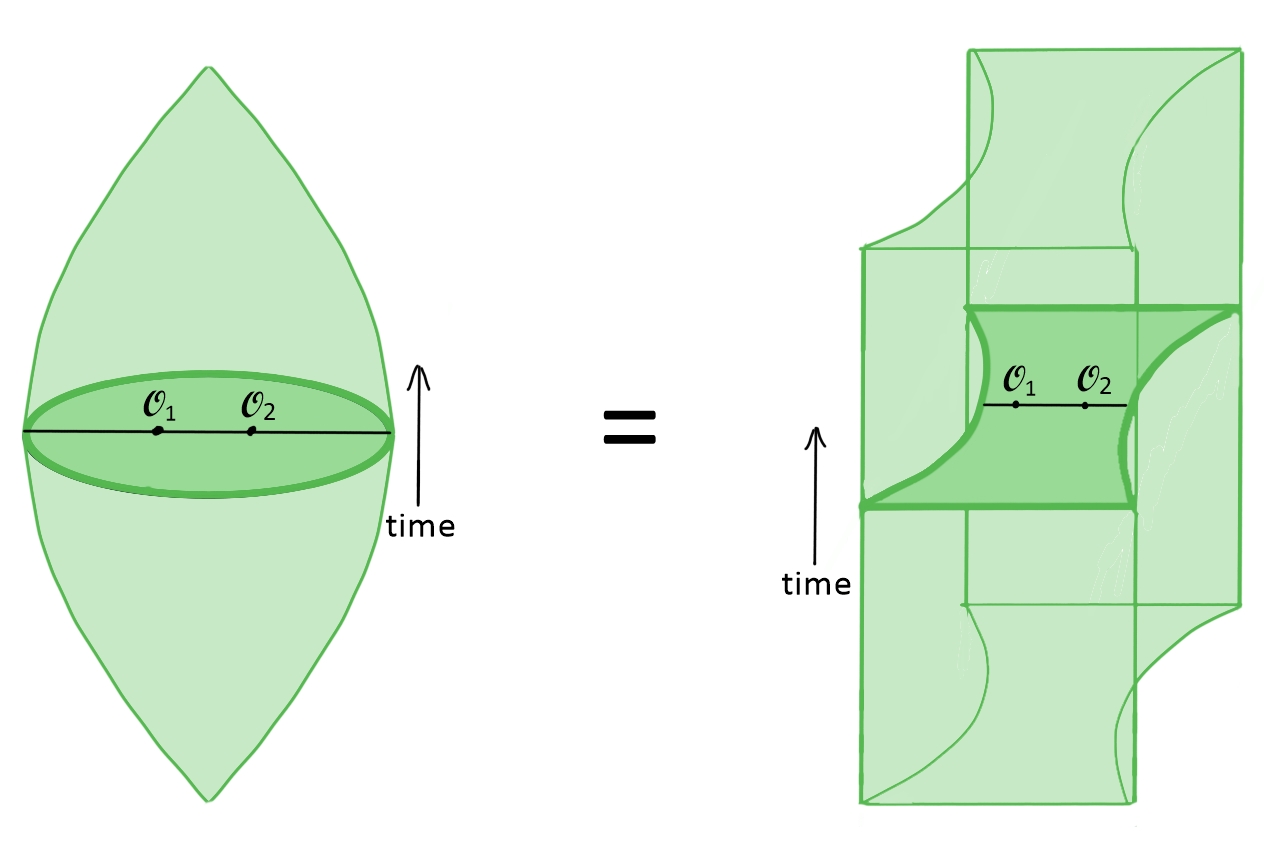}
  \caption{Observables restricted to an $\mathbb{R}^2$ at the time-reflection symmetric surface in the cosmology (black line) are equal to corresponding observables on the $\mathbb{R}^2$ in the middle of the Lorentzian wormhole. More general observables are related by double analytic continuation.}
\label{fig:operator}
\end{figure}

However, because we are considering a special state that arises from a Euclidean path integral, observables in the Lorentzian cosmology can be extracted in a simple way from this Euclidean theory, or alternatively from the vacuum physics of the Lorentzian wormhole. Since both the cosmology and the Lorentzian wormhole arise from the same Euclidean path integral through different slicings, there is a precise relationship between observables in the cosmology picture and observables in the wormhole picture. We have already seen that the scale factors in these two pictures are related by analytic continuation. But the correlation functions of local operators in the cosmology (which include all standard cosmological observables) will also be related by analytic continuation to the correlation functions in the wormhole picture. The wormhole geometry is static and horizon-free, so local observables in this picture are related in a much simpler way (e.g. by an HKLL procedure \cite{Hamilton:2005ju}) to the dual quantum field theory, which is in its vacuum state. Thus, we can understand cosmological observables by calculating the {\it vacuum} observables in the wormhole picture (via effective field theory or using the vacuum observables of the dual QFT), and then analytically continuing these to the cosmology picture. As a special case, correlation functions in the cosmology at $t=0$ (the turn-around point) in which all operators live in a single 2D plane are simply equal to some corresponding vacuum correlation functions at the midpoint of the wormhole, as shown in Figure \ref{fig:operator}.

In the wormhole picture, there is no big bang or big crunch, and the spacetime is weakly curved everywhere. Thus, an effective field theory description should be valid everywhere. This should permit a computation of cosmological observables without having to explicitly understand physics in the vicinity of the Big Bang, and without having to use the underlying microscopic description.

\subsubsection*{Naturalness problems in cosmology}

If this framework can support realistic models of cosmology it may help resolve various naturalness problems. Extrapolating our present observations of the universe back to early times we find that the universe at early times must have been extremely flat and extremely homogeneous, and must have contained correlations between regions that were apparently never in causal contact since the big bang. Thus, the required initial state would appear to be finely tuned and/or unnatural in various ways. Inflation seeks to explain how such an initial state could have arisen naturally by postulating an earlier period of exponential expansion with many e-foldings. 

In our framework, the Euclidean theory constructs a state at the time-symmetric point, and the initial state close to the big bang is most naturally understood as the result of quantum evolution backwards to the big bang.\footnote{We can consider the quantum evolution of the state with either direction of time. We discuss below in Section \ref{sec:arrow} how the arrow of time experienced by observers might emerge.} Thus, features of the state in the early universe are natural provided that they arise from the backward evolution of a  state at the time-symmetric point that is produced by the Euclidean path integral. 

The construction of the state via a Euclidean theory with $\mathbb{R}^3$ symmetry naturally produces a universe that is flat and very homogeneous. Also, since correlations in the cosmology picture at the time-symmetric slice are equal to vacuum correlators in the wormhole picture, we {\it naturally have correlations in the cosmology picture between regions of the universe that were never in causal contact}, since all points in the static wormhole picture have intersecting past light cones. 

While the model appears to alleviate many of the problems that inflation was introduced to explain, it remains to be seen whether it can reproduce the detailed quantitative results for the cosmological perturbations that arise from inflationary models and that agree with observations. An interesting point is that inflationary physics may still be present in our framework and may provide a dual explanation for the origin of perturbations/correlations, but since we can also compute these via the wormhole picture, {\it no explicit knowledge of the inflationary potential or any other early universe physics is required to compute the perturbations}.

\subsubsection*{Generic cosmological predictions}

In this paper, we focus on the generic predictions of this class of models. In all cases, the underlying $\mathbb{R}^3$ symmetry gives rise to a cosmology that is flat, homogeneous, and isotropic. The background is always time-symmetric so that the universe will eventually recollapse. For models based on 3D CFTs with relevant operators, we generically have time-dependent scalar fields whose potential energy decreases from a positive value to a negative value before rising to a larger negative value at the time-symmetric point (so we have time dependent dark energy). In some cases, the positive potential {\it leads to a phase of accelerated expansion before the collapse}. The models could have inflation in the early universe, but it is not clear whether this is generic. However, as we emphasized above, direct knowledge of the physics of the early universe is not needed to extract the physics of cosmological perturbations from the model, and these perturbations naturally contain correlations at all scales. Further, the standard flatness and horizon problems (explaining why the universe is so flat and why the temperature is nearly the same in all directions) are explained by the symmetries of the model and the fact that the state of the cosmological spacetime is naturally defined at the time-symmetric point rather than at the big bang. Thus the setup may solve the standard problems that inflation addresses without the need for inflation. However, it remains to be seen whether the possibilities for the scale factor evolution and the details of the cosmological perturbations are realistic.

The specific predictions of the model will depend on the 4D effective field theory, which is in turn determined by our choice of 3D and 4D QFTs in the microscopic setup. This choice is the only input for the physics of the model. The curvature scale in the universe is related to the number of degrees of freedom in the 3D holographic CFT\footnote{As we explain below, avoiding large quantum corrections to the cosmological constant in the effective field theory picture may require that the underlying theories are supersymmetric, but this supersymmetry may not be apparent in the 4D effective description of the cosmology because it is broken by the state, e.g. by time-dependent scalar field expectation values.}, and the gauge group in the 4D effective theory is related to the global symmetry group of the 3D CFT. Some of this gauge symmetry (and supersymmetry) may be broken by the time-dependent scalars, whose values would set the scale of the symmetry breaking. 

\subsubsection*{Outline}

In the following sections, we flesh out these basic observations and explore their consequences. In section 2, we review the basic setup and motivations for this class of models and explain how the cosmological observables are related to vacuum observables in the dual wormhole picture. In section 3, we explore the general properties and predictions for the background cosmology, making explicit the relation between cosmological observables and vacuum physics of the wormhole. We show that the simplest toy model gives rise to a standard $\Lambda < 0$ plus radiation cosmology with a big bang and a big crunch, but that more generic models (described in detail in section 3.6) will include time-dependent scalar fields that can naturally yield a phase of accelerated expansion before the collapse. In section 4, we discuss fluctuations about the background geometry and how cosmological perturbations (including physics of the CMB) are related to vacuum correlators in the wormhole. We conclude in section 5 with a discussion, including the steps necessary to verify that fully microscopic models of this scenario can work. Even if the models we consider turn out not to be viable to describe realistic cosmology, they have a number of interesting general features that might be employed in future microscopic models of cosmology based on string theory and/or holography. We summarize these features in section 5.3 of the discussion. 

\subsubsection*{Relation to earlier work}

This paper represents the present stage of evolution of a class of cosmological models originally proposed in \cite{Cooper2018}, and further developed in \cite{Antonini2019,Antonini:2021xar,VanRaamsdonk:2020tlr, VanRaamsdonk:2021qgv} (see also \cite{Wang:2021xih,Fan:2021eee}). In the original models, the 4D CFT was assumed to be holographic, and the cosmological physics was on an ETW brane behind the horizon of a black hole.\footnote{These models built on the AdS/BCFT proposal, see \cite{Karch:2001cw,Takayanagi2011,Fujita2011}, see also \cite{Kourkoulou:2017zaj, Almheiri:2018xdw}.} The present manifestation of the models with specific suggestions for microscopic realizations appeared in \cite{VanRaamsdonk:2020tlr, VanRaamsdonk:2021qgv}. The present paper is the first to explicitly consider in detail the phenomenological consequences and point out that examples making use of 3D CFTs with relevant operators can lead to accelerated expansion. 

Related models of cosmology for low dimensions were discussed in \cite{Penington:2019kki, Dong:2020uxp, Chen:2020tes}. The ``holographic cosmology'' of \cite{McFadden:2009fg} has similarities to our model in that they propose to relate cosmological physics to geometry associated with a 3D RG flow. However in that case, the transformation relating the geometries involves an analytic continuations of parameters as well as coordinates, and it is not clear whether the cosmology picture can be directly associated with an underlying physical theory. The paper \cite{hartle2012accelerated} discussed a different approach to relating accelerated expansion to $\Lambda < 0$ physics. The recent model of \cite{boyle2018c} also assumes a time-symmetric universe with an associated special state, but in that case, the point of time-reflection symmetry is the big bang, and $\Lambda > 0$ is considered. 

\section{Basic setup}
\label{sec:basic}

In this section, we explain our basic holographic construction, focusing on the physically relevant case of flat $\Lambda < 0$ four-dimensional cosmology.  

Our starting point is a three-dimensional holographic superconformal\footnote{We could also start with a non-supersymmetric conformal field theory if there exists an example dual to a spacetime with small $L_{AdS}^2/G$ and small extra dimensions. However, as we discuss below, supersymmetry at this stage may be required in order to avoid a cosmological constant problem. We will end up breaking supersymmetry in any case.} field theory on $\mathbb{R}^{2,1}$ dual to gravity on an asymptotically AdS$_4$ spacetime. For a realistic cosmology, we want to choose an example with many degrees of freedom so that the curvature scale very small relative to the Planck scale, i.e. $L_{AdS}^2/G \ll 1$. We would also like the size of the extra dimensions to be small.\footnote{The recent papers \cite{Demirtas:2021ote,Demirtas:2021nlu} provide an explicit construction of supersymmetric AdS string vacua with negative cosmological constant whose magnitude is less than $10^{-123}$ in Planck units and whose extra dimensions are very small. These 3D SCFTs dual to these solutions may be an appropriate starting point for our construction.} We can choose the global symmetry group ${\cal G}$ of the SCFT to correspond to some desired gauge group for the matter in the four-dimensional gravitational theory.\footnote{As we discuss below, this gauge symmetry may be further broken by time-dependent scalar field expectation values in the cosmology.} 

\subsubsection*{The Lorentzian wormhole and its dual}

The next step promotes the asymptotically AdS$_4$ geometry to a Lorentzian traversable wormhole with two asymptotically AdS$_4$ regions. To do this, we consider a second copy of the 3D SCFT, but with reversed orientation, so that the new copy preserves a complementary set of supersymmetries \cite{VanRaamsdonk:2021qgv}. This is dual to gravity on a second asymptotically AdS$_4$ spacetime. To connect these two asymptotically AdS spacetimes, we couple the two three-dimensional theories weakly, via an intermediate four-dimensional CFT with relatively few degrees of freedom \cite{VanRaamsdonk:2020tlr}, as in Figure \ref{fig:coupling}. The coupling introduces a scale (the distance between the 3D CFTs), and in the examples we require, the theory confines at lengths much larger than this scale. The vacuum state has a large amount of entanglement between the two 3D CFTs, and this leads to the connected wormhole geometry in the gravity dual.\footnote{See \cite{VanRaamsdonk:2020tlr,VanRaamsdonk:2021qgv} for a detailed discussion of this claim, and \cite{Betzios:2021fnm} for more recent discussion of the idea.} Models with the correct vacuum structure will exhibit a symmetry breaking ${\cal G} \times {\cal G} \to {\cal G}$ corresponding to the fact that the bulk gauge field in one asymptotically AdS region is the same as the one in the other asymptotically AdS region \cite{Antonyan:2006pg}. The IR field theory will have massless Goldstone bosons associated with this symmetry breaking.

The main evidence that this 3D-4D-3D field theory construction can give a holographic description of the Lorentzian wormhole is \cite{VanRaamsdonk:2020tlr, VanRaamsdonk:2021qgv}:
\begin{enumerate}
\item
In examples where the 4D theory is also holographic, the dual geometry has the interpretation of a geometrical brane and its corresponding anti-brane separated at the asymptotic boundary of a 5D spacetime. Parallel brane-antibrane systems are typically unstable, and the natural endpoint of the instability in this case is for the branes to connect up, forming the wormhole. We expect that the basic field theory physics responsible for this effect continues to apply even when the four-dimensional CFT is not holographic.
\item
By coupling the two 3D CFTs with a 4D CFT, we break conformal invariance and supersymmetry. The resulting theory will have an RG flow, and it is natural that the degrees of freedom in the two 3D CFTs will become more and more entangled as we go to the IR. It is also generic that the IR theory should be confining rather than having a non-trivial IR fixed point. The wormhole dual is consistent both with the large amount of entanglement between the two 3D theories and with the confinement (since the middle of the wormhole gives an IR end to the spacetime at a finite proper distance from any point in the geometry).
\item
As we review below, the effective field theory description of the wormhole requires an unusual quantum field theory phenomenon (anomalously large negative Casimir energies) that had not previously been understood. Motivated by the models, this phenomenon was searched for and found to exist in a holographic quantum field theory setup \cite{VanRaamsdonk:2021qgv, May:2021xhz}. 
\item
The basic mechanism of coupling two holographic CFTs to generate a traversable wormhole is similar to that used in \cite{Gao2016, Maldacena:2018lmt}, though our mechanism crucially involves auxiliary degrees of freedom to achieve the coupling.
\end{enumerate}

Going forward, we will assume that this holographic description is correct; in the discussion we comment on ways that this could be verified explicitly in microscopic examples.

\subsubsection*{The Euclidean wormhole and its dual}

So far, we have the vacuum state of a 3D-4D-3D QFT dual to a geometry with a four-dimensional planar Lorentzian wormhole. The wormhole geometry can be described using a metric
\be
\label{worm_proper_time}
ds^2 = d\tau^2 + A^2(\tau) (-d \zeta^2 + dx_i dx_i) \; .
\ee
Here, $A(\tau)$ is symmetric under reversal of $\tau$ and grows exponentially for $\tau \to \pm \infty$ corresponding to the two asymptotically AdS boundaries. The $\zeta$ and $x^i$ coordinates are the coordinates of the original 3D CFT. 

The vacuum state of our Lorentzian field theory is naturally associated with a Euclidean version of this theory, where the 3D CFTs are on $\mathbb{R}^3$ and the 4D theory is on $\mathbb{R}^3$ times an interval $I$. Specifically, the vacuum wavefunctional is obtained by slicing the path integral defining the Euclidean theory perpendicular to one of the $\mathbb{R}^3$ directions, as reviewed in appendix A.

The Euclidean theory is dual to the analytically continued wormhole geometry, with metric:
\be
ds^2 = d\tau^2 + A^2(\tau) (dw^2 + dx_i dx_i) \; .
\ee
More precisely, this is the saddle point geometry for the gravitational path integral of the dual theory.

\subsubsection*{The cosmology and its dual}

The cosmological spacetime that we desire is obtained from this Euclidean wormhole spacetime by analytic continuation of the $\tau$ coordinate, which gives the planar FRW geometry
\be
\label{FRW}
ds^2 = -dt^2 + a^2(t) (dw^2 + dx_i dx_i) \; .
\ee
In this case, the function $a(t)$ is time-symmetric and will generally correspond to a big-bang / big-crunch cosmology, though other time-symmetric scale factors might be possible.\footnote{See \cite{Antonini2019} for a possible construction of bouncing cosmology.}

The holographic construction gives rise to a special state describing the quantum physics of this cosmology. This is obtained by slicing the path integral for the Euclidean theory in a different way, at the midpoint of the interval $I$, as shown in Figure \ref{fig:slices}. This new slicing does not intersect the 3D CFTs at all, so the interpretation is that we have an excited state of the 4D CFT on $\mathbb{R}^3$ produced by a path integral that has the ``insertion'' of a 3D holographic CFT in the Euclidean past.

Unlike usual examples of holography, the underlying degrees of freedom are not associated with any boundary of the cosmological spacetime. We will understand this better presently.

\subsection{Higher dimensional description with a 4D holographic CFT}

To develop intuition for this class of models, it is useful to understand a special case, where the 4D auxiliary theory is also holographic. This was the case considered in \cite{Cooper2018, Antonini2019} where such models were originally proposed. 

Here, each geometry described above is now the geometry of an end-of-the-world (ETW) brane that lives at the boundary of a five-dimensional spacetime (bottom row of Figure \ref{fig:NinePictures}). 

For the Lorentzian and Euclidean wormhole geometries, the five-dimensional geometry is asymptotically AdS with boundary geometry $\mathbb{R}^{2,1} \times I$ or $\mathbb{R}^{3} \times I$, and the ETW brane reaches the asymptotically AdS region at the boundaries of $I$. 
In the cosmology picture, the higher-dimensional geometry is asymptotically AdS with boundary geometry $\mathbb{R}^{3,1}$, and the ETW brane does not intersect the boundary. The dual state is a high-energy excited state of the 4D CFT, so the geometry can be interpreted as a particular planar black hole microstate. The ETW brane housing the FRW geometry lives behind the horizon of this black hole. The cosmological singularities of the FRW spacetime are part of the past and future singularities of the black hole. 

This last picture provides intuition for how the 4D cosmological spacetime can be described by some state of the 4D CFT which is not associated with any boundary in the cosmology. From the 5D point of view, the cosmology and the asymptotically AdS boundary are separated in the 5D radial direction. In examples where the 4D theory is not holographic, there is no 5D geometrical space, so the cosmology is like an island (using the terminology of \cite{Maldacena2013,VanRaamsdonk:2013sza,Penington:2019npb,Almheiri2019b}). Other examples where a holographic theory in the Euclidean past is used to encode a geometrical spacetime in a quantum theory that is not necessarily holographic were discussed recently in \cite{Simidzija:2020ukv} and \cite{VanRaamsdonk2018}. This is also similar to the way that black hole interiors can be encoded in auxiliary radiation systems \cite{Maldacena2013, VanRaamsdonk:2013sza, Penington:2019npb, Almheiri2019b}.

\subsection{Auxiliary systems in the effective field theory description}

We now return to the generic case to better understand the effective field theory description in the cosmology, wormhole, and Euclidean pictures. We will see that the 4D CFT plays an important role as a non-gravitational auxiliary system in the effective descriptions. 

\paragraph{Single 3D theory:}  The 3D holographic CFT that we start with is dual to some gravitational theory whose 4D effective description has gravity with a negative cosmological constant and matter with gauge symmetry given by the global symmetry group of the 3D CFT. 

\begin{figure}
  \centering
  \includegraphics[scale=0.32]{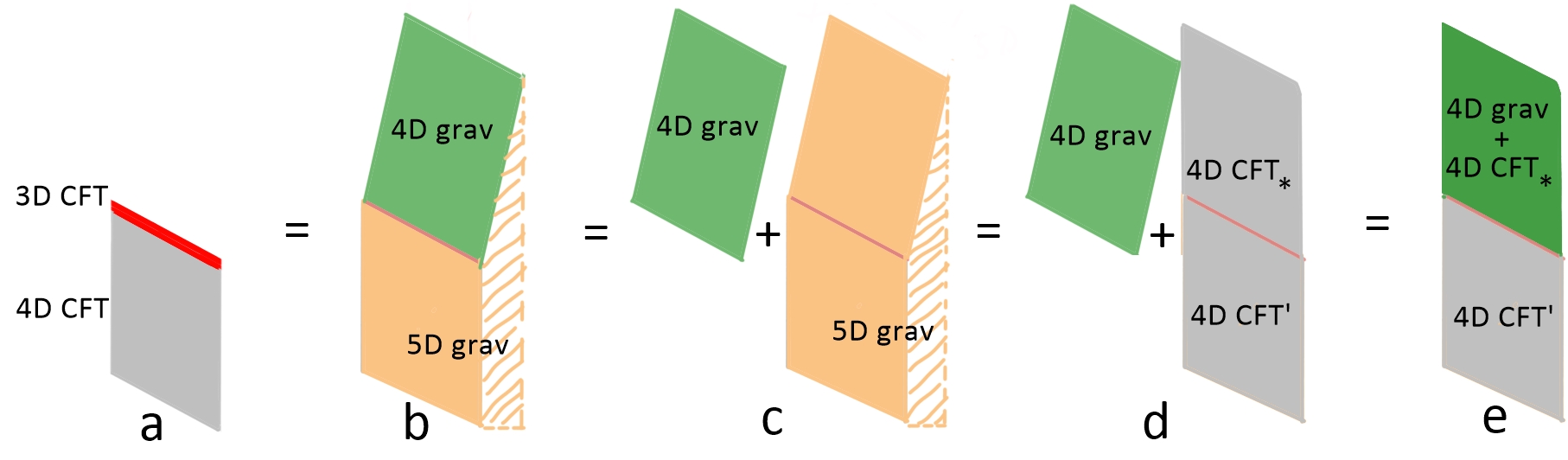}
  \caption{Effective field theory description of holographic 3D-4D coupled theories with $c_{3D}\gg c_{4D}$. a) Microscopic picture b) Higher-dimensional gravity picture: 4D gravitational physics of the 3D theory appears as the physics of an ETW brane near the AdS boundary. c) Isolating the 5D bulk gravitational physics. d) 5D gravitational physics is equivalent to physics of the dual 4D CFT with a cutoff (indicated by *) on half the space. e) Final effective description: a 4D CFT coupled to a 4D gravity theory + cutoff CFT. The state of the lower 4D CFT in the final picture is different than the state in the micoscopic picture.}
\label{fig:EFT}
\end{figure}

\paragraph{Single 3D theory + 4D theory:} Next, we would like to understand the effect of coupling a single copy of the 3D CFT to a 4D CFT on a half-space. 

This can be understood most easily by thinking about the case where the 4D CFT is holographic (see Figure \ref{fig:EFT}). Here, we have a higher-dimensional dual description where the 4D gravity theory dual to the 3D CFT now gives the effective description of physics on an ETW brane in a 5D asymptotically AdS spacetime whose boundary geometry is a half space.\footnote{Such models were considered originally in \cite{Karch:2000ct,Karch:2000gx,Karch:2001cw}. Microscopic examples starting with ${\cal N}=4$ SYM theory were discussed in \cite{Gaiotto:2008ak}, and explicit gravity dual solutions were found in \cite{DHoker:2007zhm, DHoker:2007hhe, Aharony:2011yc, Assel:2011xz}, based on the general $OSp(2, 2|4)$-symmetric solutions of type IIB supergravity found in \cite{DHoker:2007zhm, DHoker:2007hhe}. The limit of $c_{3D} \gg c_{4D}$ was considered in \cite{Bachas:2018zmb,VanRaamsdonk:2021duo,Uhlemann:2021nhu}. } In the limit of interest where the 3D theory has many more local degrees of freedom than the 4D CFT, the ETW brane makes an angle close to $\pi/2$ with the bulk radial direction and lies on a surface that cuts off the would-be second half of the asymptotic region.

From the higher-dimensional perspective, the extra physics that we have added via the 4D CFT is a gravitational theory on asymptotically AdS spacetime with a cutoff surface that removes one half of the asymptotic region. In field theory language, this should correspond to the 4D CFT on Minkowski space with a UV cutoff on half of the spacetime.\footnote{Here, ``cutoff'' really indicates some UV modification that allows the theory to be coupled in to the gravitational theory in way that preserves supersymmetry.} In this half of the spacetime the cutoff 4D CFT is coupled locally to the 4D gravitational theory dual to the 3D CFT (Figure \ref{fig:EFT}e).\footnote{According to \cite{Porrati:2001gx}, this coupling can also be understood as a modification to the gravitational effective action, in which the graviton develops a small mass, related to the ratio $c_{4D}/c_{3D}$ of degrees of freedom, and certain higher curvature terms are induced.} While we have motivated this final picture by higher-dimensional holography, we expect it to hold also when the 4D CFT is not holographic.

\begin{figure}
  \centering
  \includegraphics[scale=0.35]{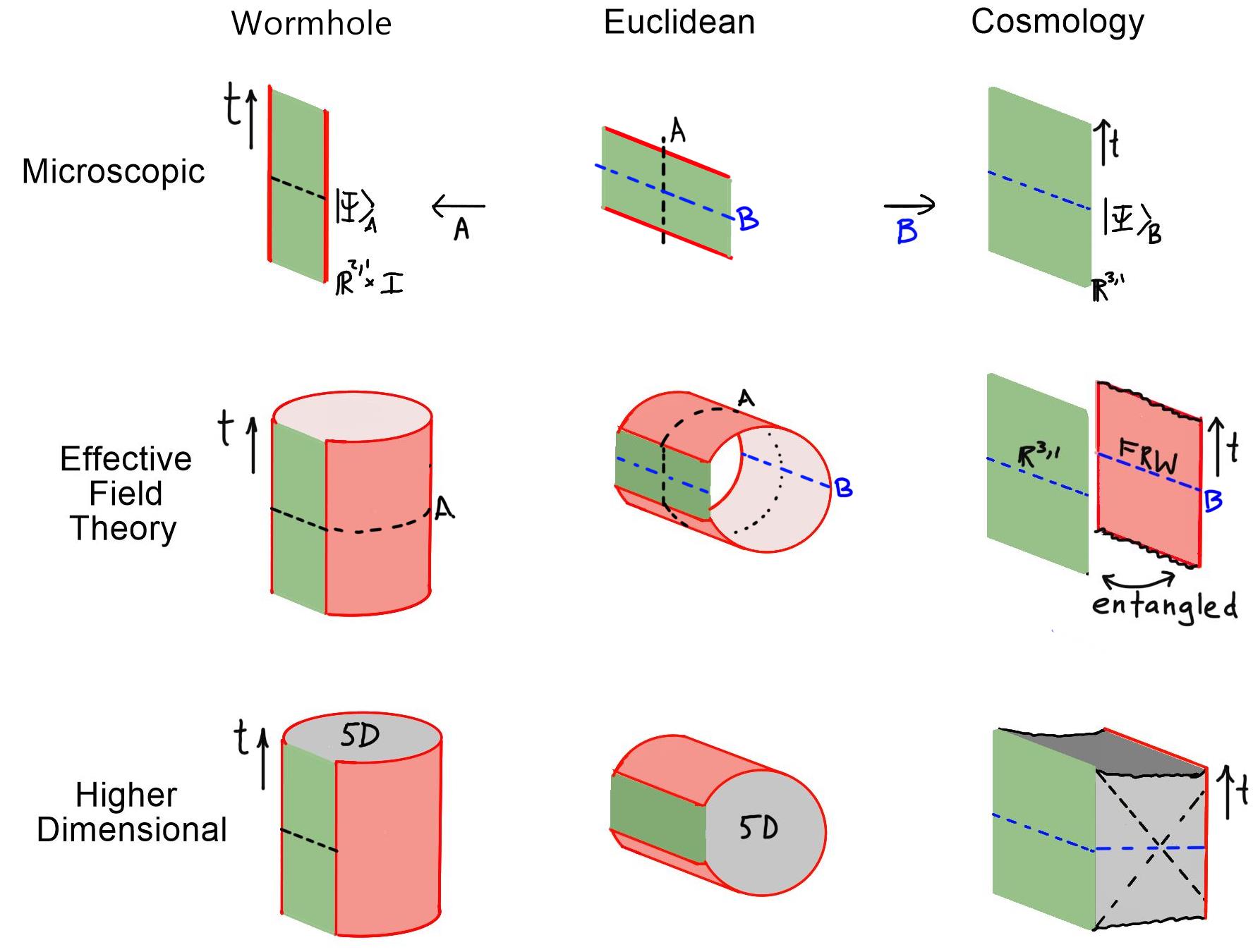}
  \caption{Top row: the Euclidean path integral for the 3D-4D-3D theory (middle-top) can be sliced in two different ways to the vacuum state of the Lorentzian 3D-4D-3D theory or a complex excited state of the 4D CFT. The middle row shows the holographic dual descriptions of these pictures in the general case where the 4D theory is not holographic. The Euclidean description (center) can be understood as a gravitational path integral that can be sliced to define the static Lorentzian wormhole or the cosmology, entangled with an auxiliary 4D CFT. When the 4D CFT is holographic, we have a higher-dimensional gravity description (bottom row).}
\label{fig:NinePictures}
\end{figure}

\paragraph{Two 3D theories + 4D theory:} In the full construction for the Lorentzian wormhole, we have two boundary 3D CFTs that are coupled together by a 4D CFT. In this case, we have two copies of the four-dimensional gravitational theory just described, each with an asymptotically AdS region, but these join up in the IR to give the wormhole geometry. The fields at the two asymptotically AdS boundaries are coupled via the auxiliary non-gravitational 4D CFT on an interval. This is depicted in the middle-left picture of Figure \ref{fig:NinePictures}.\footnote{This discussion has been somewhat heuristic. It would be nice to understand better the extent to which the effects of the higher-dimensional bulk geometry can be captured by a local effective field theory. The discussion below assumes that any effects that are non-local from the four-dimensional point of view can be neglected.} 

This auxiliary 4D theory coupling the two asymptotically AdS boundaries is an essential part of the construction. Without it, the vacuum state of the quantum fields on the wormhole would be significantly different and would not exhibit the type of vacuum energy to support the wormhole in the first place. We will discuss this further in section 3.

\paragraph{Euclidean picture:} The effective field theory description of the Euclidean picture is essentially the same as that in the Lorentzian wormhole picture, but with the translation-invariant time direction replaced by a spatial direction (Figure \ref{fig:NinePictures}, middle). 

\paragraph{Cosmology picture:} The Euclidean effective field theory (Figure \ref{fig:NinePictures}, middle) has a reflection symmetry in the direction between the two asymptotically AdS boundaries of the wormhole. The surface fixed by this reflection (labeled B in the center picture of Figure \ref{fig:NinePictures}) has two disconnected parts, one in the gravitational theory at the midpoint of the wormhole, and one in the 4D CFT. This surface corresponds to the $t=0$ surface in the cosmology picture, so that picture also includes a gravitational theory (housing the cosmology) and an auxiliary 4D CFT. These do not interact with each other directly, but are entangled. The connection between these two parts in the lower half of the Euclidean picture leads to entanglement between the two parts in the cosmology picture, similar to how a thermofield double path integral connects two quantum systems in the Euclidean past.

When we have a 5D gravity description, we can think of the state of the 4D CFT as describing the physics outside the black hole horizon and the 4D gravitational theory as describing the physics inside the black hole including the physics of the ETW brane. 

The role of the 4D CFT in the effective description (Figure \ref{fig:NinePictures}, middle-right) is different from its role in the microscopic description (Figure \ref{fig:NinePictures}, top-right). In the micsrocopic picture of the cosmology, we only have a complicated pure state of the 4D CFT. The physics of the ETW brane is encoded in the IR physics of this state. In the effective field theory picture, we have a simpler mixed state of the 4D CFT that encodes the physics outside the horizon, and this is purified by the state of the 4D gravitational theory that describes the cosmological physics. As we describe in the discussion, this implies the presence of ``islands'' and explains the appearence of cosmological islands in the work of \cite{Hartman:2020khs}.

As summarized in Figure \ref{fig:NinePictures}, the different analytic continuations, together with the holographic duality and the possibility of a higher-dimensional holographic description give up to nine different pictures of the physics.

\subsection{Cosmology from vacuum physics}

In our setup, we have two Lorentzian spacetimes, the cosmology and the Lorentzian wormhole, whose holographic descriptions arise from slicing the same Euclidean path integral in two different ways. This ``path integral slicing duality'' implies an exact equivalence between certain observables for the two Lorentzian states. The Euclidean path integral with operators inserted on the intersection of the two slices will have an interpretation as the expectation value of some observables in one Lorentzian picture and some closely related observables in the other Lorentzian picture; these are equal since they are computed in exactly the same way. 

In our application, the Euclidean path integral in the 3D-4D-3D QFT corresponds to a Euclidean gravitational path integral that includes contributions from a dominant saddle point geometry (the Euclidean wormhole) and fluctuations around this. The geometry and fluctuations here determine the background geometry and quantum states in the two Lorentzian pictures. To the extent that we can describe the Lorentzian physics in terms of quantum field theory on a background saddle-point geometry, we can say that the field theory observables in the cosmology picture will be directly related to field theory observables in the wormhole picture, since again they arise from the same Euclidean path integral.\footnote{Going beyond the field theory approximation, we run into the question of how to precisely define bulk gravitational observables; it may be that for the precisely defined gravitational observables associated to a naive local observables in the QFT picture, we have slight differences between the observables in the wormhole and the cosmology picture.} 

The simplest case is to consider some observables that live at the intersection of the two slices. For example, we can consider in the Euclidean effective field theory picture some correlation function of local operators restricted to an $\mathbb{R}^2$ in the $\mathbb{R}^3$ that forms the midpoint of the wormhole. This Euclidean correlation function is equal to some correlation function in the cosmology at the time-reflection symmetric surface, and also equal to a vacuum correlation function of operators at the midpoint of the wormhole in the Lorentzian wormhole picture. In this way, any cosmological observable that lives in an $\mathbb{R}^2$ at the time-symmetic point is equal to a vacuum observable in the wormhole picture, as shown in Figure \ref{fig:operator}.

More generally, time-dependent observables in each Lorentzian theory are related by analytic continuation to observables in the Euclidean theory, so time-dependent observables in the two Lorentzian theories will be related by a double analytic continuation. We have already seen that this applies to the scale factor in the background geometry. But it should also apply to other observables, such as the equal time correlation functions of the stress-energy tensor that encode the physics of the CMB. Thus, by understanding the vacuum observables in the Lorentzian wormhole picture, we should be able to answer almost any question about the physics of the cosmology.

In the remainder of this paper, we focus on the effective field theory description, and the relation between the cosmological observables in the FRW universe and vacuum observables in the wormhole geometry dual to the vacuum state of the 3D-4D-3D theory.

\section{Implications for background cosmology}

In this section, we focus on the background cosmology, describing how the scale factor evolution and stress-energy tensor evolution are related in the cosmology and wormhole pictures. 

Assuming the validity of four-dimensional effective field theory, we expect that it should be a good approximation when discussing the background geometry to use the semiclassical Einstein equation,
\be
\label{semiclassical}
R_{\mu \nu} - {1 \over 2} g_{\mu \nu} R - {3 \over L^2} g_{\mu \nu} = 8 \pi G \langle T_{\mu \nu} \rangle \; ,
\ee
where $L$ is the AdS length associated with the underlying 3D CFTs. It will be useful to begin by explicitly describing the background geometries and evolution equations in the two pictures.

\subsubsection*{Background geometry and stress-energy tensor: cosmology}

In the cosmology picture, we have a background geometry
\be
\label{FRW1}
ds^2 = -dt^2 + a^2(t) (dw^2 + dx_i dx_i) \; ,
\ee
where $a(t)$ is time-reversal symmetric.

The homogeneous part of the stress-energy tensor takes the form\footnote{Here and below, quantum expectation values are implied when writing components of the stress-energy tensor.}
\be
T^{tt} = \rho(t) \qquad T^{ij} = {1 \over a^2(t)} \delta^{ij} p(t) \qquad T^{ww} = {1 \over a^2(t)} p(t) \; .
\ee
where $\rho$ and $p$ are the conventionally defined energy density and pressure. These will also be time-reversal symmetric.
In terms of the Hubble parameter $H = \dot{a}/a$, the covariant conservation equation $\nabla_\mu T^{\mu \nu} = 0$ gives
\be
\label{Econs1}
\dot{\rho} = -3 H (\rho + p) \;
\ee
which can be used to express $p$ in terms of $\rho$ and $a$. Finally, Einstein's equations relate $\rho$ and $a$ by the Friedmann equation
\be
\label{Friedmann}
H^2 + {1 \over L^2} = {8 \pi G \over 3} \rho \; .
\ee
Taking the time derivative of (\ref{Friedmann}) and using (\ref{Econs1}), we have also that
\be
\dot{H} = - 4 \pi G (p + \rho) \; .
\ee
In this cosmology picture, we expect that the stress-energy tensor has various contributions, including vacuum energy of fields as well as matter and radiation.

\subsubsection*{Background geometry and stress-energy tensor: wormhole}

Let us now see how the scale factor and stress-energy tensor in the cosmology are related directly to those in the Lorentzian wormhole spacetime. 

In the wormhole picture, we can describe the metric as
\be
\label{worm}
ds^2 = d\tau^2 + a_E^2(\tau) (-d \zeta^2 + dx_i dx_i) \; .
\ee
Here, $a_E(\tau)$ is obtained from $a(t)$ by analytic continuation $t^2 \leftrightarrow -\tau^2$. It is symmetric under reversal of $\tau$ and grows exponentially for $\tau \to \pm \infty$ corresponding to the two asymptotically AdS boundaries.

In this picture, the stress-energy tensor is 
\be
T^{\tau \tau} = -\rho_E(\tau) \qquad T^{ij} = {1 \over a^2(t)} \delta^{ij} p_E(\tau) \qquad T^{\zeta \zeta} = -{1 \over a_E^2} p_E(\tau) \; ,
\ee
where $\rho_E$ and $p_E$ are related to $\rho$ and $p$ by analytic continuation $t^2 \leftrightarrow -\tau^2$.

The energy conservation equation gives
\be
\label{Econs}
{d \rho_E \over d \tau} = -{3 \over a_E} {d a_E \over d \tau}  (\rho_E + p_E) \;
\ee
and the analogue of the Friedmann equation gives
\be
\label{FriedmannE}
-\left( {1 \over a_E} {d a_E \over d \tau} \right)^2 + {1 \over L^2} = {8 \pi G \over 3} \rho_E \; .
\ee

\subsection{Energy from vacuum energy}

We see that the time evolution $\rho(t)$ of the cosmological energy density, is directly related to a component of the vacuum stress tensor in the wormhole picture as 
\be
\rho(t) = - T_{\tau \tau}(\tau = it)
\ee
where the right side is real because of the $\tau \to - \tau$ symmetry. 

The pressure in the cosmology picture is \be
p(t) = T^\zeta {}_\zeta(\tau = it) \; ,
\ee
where the right side is minus the proper energy density in the wormhole picture. Here, $p(t)$ is also real by the $\tau$-reflection symmetry. While $T_{\tau \tau}$ and $T^\zeta {}_\zeta$ represent components of the vacuum stress energy tensor in the wormhole picture (i.e. a Casimir stress-energy tensor in the system with two boundaries), the functions $\rho(t)$ and $p(t)$ represent the full stress-energy tensor in the cosmology, including both vacuum and non-vacuum contributions. Thus, in a realistic example, the stress-energy tensor from matter, dark matter, and vacuum in the cosmology picture would all arise from vacuum energy in the wormhole picture. They could be computed from knowledge of the effective field theory in the wormhole and are thus determined without specific knowledge of the big bang or the initial state in the cosmology picture.

\subsection{Analytic structure of the scale factor}

The $\tau \to - \tau$  reflection symmetry and AdS asymptotics in the wormhole picture imply that the wormhole scale factor $a_E(\tau)$ is an even function of $\tau$ (defining $\tau = 0$ to be the middle of the wormhole) that grows exponentially for $\tau \to \pm \infty$.\footnote{Alternatively, the scale factor $\hat{a}_E(z)$ in conformal coordinates (see appendix B) has simple poles at $\pm z_0$, with no other poles in $[-z_0,z_0]$.} 

The Lorentzian scale factor $a(t)$ is obtained from $a_E(\tau)$ by analytic continuation $\tau \to it$, so it will also be an even function of $t$. That $a_E(\tau)$ has the asymptotics of a $\cosh$ function suggests an oscillatory piece in the real time scale factor $a(t)$ (since $\cosh$ continues to a cosine), but this does not necessarily imply a periodic cosmology. For example, in the simple example $a_E^2(\tau) = \cosh(2\tau/L)$ that arises from a $\Lambda < 0$ + radiation cosmology, we find $a^2(t) = \cos(2t/L)$ which goes to zero for $t = \pm \pi/(2L)$. Thus, we get a big-bang/big-crunch.

For other choices of $\hat{a}_E(z)$ with AdS asymptotics, we can have big-bang/big-crunch scale factors with different functional form or periodic scale factors. However, the physically allowed scale factors will be determined by the dynamics of the theory.

\subsection{Vacuum energy in the wormhole picture}

To understand the possibilities for the scale factor and stress-energy tensor evolution in the cosmology picture, we need to understand the possibilities for the vacuum energy in the wormhole picture. 

Here, it is important that the metric is dynamical, so one needs to understand the vacuum configuration of a dynamical metric coupled to quantum fields (including the metric itself).

Assuming the validity of four-dimensional effective field theory, we expect that it should be a good approximation to self-consistently solve the semiclassical Einstein equation (\ref{semiclassical}) where $\langle T_{\mu \nu} \rangle$ represents the vacuum energy of the fields (presumably, including metric fluctuations) in our effective theory on the background with metric $g$.\footnote{It is probably useful to keep in mind the alternative possibility that the four-dimensional effective field theory is not adequate to understand the background geometry, e.g. because the effects of the higher-dimensional bulk cannot be adequately reproduced by local terms.} That is, we find the vacuum energy of the effective field theory on a fixed geometry of the form (\ref{worm}), giving some functional of the scale factor $\hat{a}_E(z)$, and then include this functional on the right-hand side of the Einstein equation. In this calculation, it is important to take into account the 4D CFT coupling fields in the two asymptotically AdS regions.

The vacuum energy calculation will generally be divergent in the continuum QFT limit, but it is natural in the gravitational context to assume a cutoff at the Planck scale, $M_{UV} = 1/\sqrt{G}$. In this case, for a renormalizable quantum field theory, the divergent pieces of the stress-energy tensor should take the form of local terms that may involve the curvature. 

\subsubsection*{The cosmological constant problem}

The leading potential divergence takes the form $\langle T_{\mu \nu} \rangle = \hat{\Lambda} g_{\mu \nu}$, giving a quantum correction to the cosmological constant. Here, $\hat{\Lambda}$ could have contributions of order $1/G^2$, $M^2/G$, or $M^4 \log(M^2 G)$ for some mass scale $M$ associated with the effective theory. Any of these would be problematic for realistic cosmology: this is the usual cosmological constant problem.\footnote{The last term could be acceptable in the case that the mass scale is similar to that expected for neutrino masses.} However, there is reason to believe that such contributions will cancel in our setup. 

An acceptably small cosmological constant / curvature scale in the 4D geometry dual to a single 3D CFT can be arranged by making an appropriate choice of CFT. Supersymmetric AdS compactifications of string theory with acceptably small curvature scale and small extra dimensions have been argued to exist recently in \cite{Demirtas:2021nlu,Demirtas:2021ote}. These should correspond to some dual superconformal theory where the smallness of the curvature / cosmological constant relative to the Planck scale is related to the large number of degrees of freedom. Evidently, any $\langle T_{\mu \nu} \rangle \propto g_{\mu \nu}$ contributions to the stress-energy tensor in the effective field theory description of the 4D bulk must cancel. This can likely be understood as a result of supersymmetry in the bulk effective theory. 

Coupling in a 4D theory with many fewer local degrees of freedom in a way that preserves supersymmetry and conformal invariance should not affect the dual geometry significantly; this is known to be true in various examples where the dual geometries are known explicitly \cite{Bachas:2018zmb}.

In the full setup for the cosmology picture, we break supersymmetry by introducing another boundary theory; this preserves supersymmetry on its own, but the two boundaries together break supersymmetry. There is a well known example where such non-local breaking of supersymmetry in the CFT does not lead to large contributions to the bulk stress-energy tensor. For a supersymmetric holographic conformal field theory compactified on a circle with antiperiodic boundary conditions for fermions, supersymmetry is broken by the boundary conditions on the circle, but the vacuum state is believed to be dual to a geometry with the same scale of curvature as in the dual of the theory with supersymmetric (periodic) boundary conditions for fermions \cite{Witten1998a}. From an effective field theory point of view, the point may be that the same cancellations that avoided large contributions to the stress-energy tensor in the supersymmetric case persist here.\footnote{A useful example may be the calculation of Casimir energy density in a supersymmetric theory where the supersymmetry is broken by boundary conditions. In general Casimir energy calculations, after introducing a regulator, the energy density for infinite volume is subtracted from the energy density with boundaries, and then the regulator is removed to yield a finite result. For a supersymmetric theory, energy density in infinite volume would vanish, so the full energy density should also be finite, even though the boundary conditions break supersymmetry.} 

An alternative argument for the absence of corrections to the cosmological constant is that in the cosmology picture, the dual theory is just the 4D CFT with unbroken supersymmetry. However, the effective theory in the cosmology also then has an underlying supersymmetry but may look non-supersymmetric because of our choice of state (e.g. due to the time-dependent scalar fields). In this case, the leading divergent contributions to the stress-energy tensor should be absent since they are independent of the state and vanish for the vacuum state of a supersymmetric theory.

\subsubsection*{Renormalization of the Newton constant}

The next potential divergence comes at order $1/G$ and is proportional to the Einstein tensor, $\langle T_{\mu \nu} \rangle \sim 1/G (R_{\mu \nu} - g_{\mu \nu} R)$. Since this is multiplied by $8 \pi G$ in the Einstein equation, it comes in at the same order as the Einstein tensor on the left side. Thus, we can think of it as renormalizing the Newton constant by an order one amount. Such contributions can occur even for supersymmetric theories. 

\subsubsection*{$R^2$ terms}

The final possibility for divergent terms are $R^2$ corrections with coefficients of order $\log(M^2 G)$. Since they come into the Einstein equations multiplied by $G$, they will only be comparable to the Einstein term in regions of the geometry where the curvature is close to the Planck scale, e.g. near the initial and final time in the cosmology picture. We expect that they can be ignored in the wormhole picture.

\subsubsection*{Finite terms}

The remaining part of the stress-energy tensor, finite in the $G \to 0$ limit, can be understood as a Casimir energy density associated with the fields. This part is essential for ending up with a wormhole geometry that can give rise to a cosmological spacetime, since the other significant terms are the Einstein term and cosmological constant term that together give rise to a single-sided pure AdS spacetime. 

If the scale of curvature is of order the AdS scale $L$ throughout the geometry, the Einstein term and the cosmological constant term in the Einstein equations will both be of order $1/L^2$. In this case, the Casimir energy density must be of order $1/(G L^2)$ to have a comparable effect. Naively, the scale of the energy density for a field theory with $c_0$ fields in a geometry with length scale $L$ would be $c_0/L^4$. The number of fields in our effective theory is expected to be relatively small, so we require that the Casimir energy density is larger than its ``natural'' value by a factor of $L^2/(G c_0) \sim c_{3D}/c_0$. Thus, we have the unusual situation of needing a vacuum energy that is anomalously {\it large}. 

A more specific quantitative connection between the Casimir energy density and the scale of curvature in the geometry is provided by evaluating the $++$ component of the Einstein equation at the middle of the wormhole where $\hat{a}'_E = 0$. For this component, the cosmological constant term does not contribute, so the connection between the curvature scale and the energy density is more direct. We have that
\be
\label{Tplusplus}
T_{++} = {1 \over 8 \pi G } R_{++} = - {1 \over 4 \pi G} {\hat{a}_E'' \over \hat{a}_E} \sim - {\ell^2 \over G} {1 \over \ell^4}\; .
\ee
Choosing coordinates where $a(0)=1$, the left side gives the proper null energy density and $\ell$ gives the length scale associated with the curvature at the midpoint of the wormhole. Thus, to get $\ell \gg \sqrt{G}$, we need negative null Casimir energy that exceeds the geometrical scale ${1 / \ell^4}$ by a large factor ${\ell^2 \over G}$.\footnote{In appendix B, we show that the integrated null-energy along a null path from one asymptotically AdS boundary to the other must also be negative. Thus, the averaged null energy condition is also violated, but this is typical for Casimir energies.}

Since such a large factor is absent for the Casimir energy density of a field on a space with a periodic spatial direction\footnote{There, the Casimir null energy density is equal to the negative of the thermal null energy for the same theory at temperature $1/\ell$, as we review in Appendix D.}, the interface in the effective field theory picture between the fields on the wormhole and the non-gravitational 4D CFT fields must play an essential role. To understand whether certain interfaces can give rise to such enhancement, \cite{VanRaamsdonk:2021qgv, May:2021xhz} studied holographic models of interface quantum field theories. Remarkably, it was found that the needed enhancement of Casimir energy appears (and, within the model studied, can be arbitrarily large) for special choices of the interface physics.\footnote{In the holographic model, the interface between CFTs is associated with a bulk interface brane that separates regions of the geometry associated with the two CFTs. The enhancement of Casimir energy occurs when the tension of this interface brane is taken to some lower critical value.} 

The somewhat surprising appearance of this energy enhancement provides some evidence for the viability of the class of models we are studying.\footnote{Specifically, the general arguments for the existence of solutions together with the assumption that the solutions could be understood within the context of four-dimensional effective field theory required/predicted a seemingly unlikely quantum field theory effect, and that effect was later found to be present in holographic models.} However, it will be important to understand better how general this enhancement is (e.g. whether it requires strongly coupled field theory) and whether we can exhibit it in microscopic quantum field theories. See appendix D for additional comments related to the understanding of enhanced Casimir energies in quantum field theory.

\subsubsection*{Classical scalar field contributions}

Another possible contribution to the consistent with the symmetries of the wormhole geometry is to have one or more scalar fields with a non-trivial classical profile $\phi(\tau)$. We will see below that having non-trivial scalars is generic when the 3D CFT that we start with has relevant or marginally relevant operators, as the variation of these scalars reflect possible RG flows that we might have with this 3D CFT in the UV \cite{freedman1999renormalization}. 

With varying scalars, potential terms and derivative terms will both contribute to the stress-energy tensor and back-react on the geometry. In the cosmology picture, we will in this case have a time-dependent scalar field; we will see below that this can give rise to a phase of accelerated expansion before the recollapse. We will discuss this in more detail in section 3.6 below.

The presence of a non-trivial scalar field does not remove the need for a large Casimir energy. For example, in the wormhole picture with conformal distance coordinate $z$, we have that the contribution of the scalar field to the null energy is 
\be
T_{++} = \partial_z \phi \partial_z \phi > 0 \; .
\ee
This gives a positive contribution to the left side of equation (\ref{Tplusplus}), so the large negative Casimir energy is still required to agree with the right side of the equation.

\subsection{Example: CFT matter}

We now describe a few simple examples of our scenario at the level of effective field theory.

To begin, we review the case where the matter in the gravitational part of our effective field theory is taken to be a CFT \cite{VanRaamsdonk:2021qgv}. This is not what is expected from the microscopic models, but it has the advantage of being analytically tractable. In particular, the form of the stress-energy tensor is determined up to an overall constant by conformal invariance. 

To find the form of stress-energy tensor, we can make use of conformal coordinates (see Appendix \ref{app:conformal}) and perform a conformal transformation to flat space $\mathbb{R}^{2,1} \times [-z_0,z_0]$. Conformal invariance and 2+1 Poincar\'e symmetry imply that the vacuum stress tensor in this conformal frame takes the form
\be
T_{zz} = - {3 \over z_0^4} F(z_0) \qquad T_{\mu \nu} = \eta_{\mu \nu} {1 \over z_0^4} F(z_0)
\ee
where $F$ is a dimensionless constant that could depend on the ratio between $z_0$ and any other parameter with dimensions of length.\footnote{This could be the width of the strip of 4D CFT that couples the two asymptotically AdS regions, or some parameter associated with the interface.}

We recall that the field theory in the wormhole geometry is coupled at its boundaries via a 4D CFT on a strip. The constant $F$ can depend on our CFT, the non-gravitational CFT that couples the fields at the boundaries of the wormhole, and the physics of the interface between the two CFTs. 

In the original conformal frame, the stress tensor using conformal coordinates becomes
\be
T_{zz} = - {1 \over \hat{a}_E^2} {3 \over z_0^4} F + T_{zz}^{CA}  \qquad T_{\mu \nu} =  \eta_{\mu \nu} {1 \over \hat{a}_E^2} {1 \over z_0^4} F  + T_{\mu \nu}^{CA}
\ee
where $T_{zz}^{CA}$ and $T_{\mu\nu}^{CA}$ are terms due to the 4D conformal anomaly. In general, these are a specific combination of curvature squared terms whose coefficients depend on our choice of CFT. As in our discussion above, we expect that these can be ignored when the curvature of the geometry is much larger than Planck scale, i.e. away from the big bang/big crunch. We will drop them for now in our discussion of the wormhole picture.\footnote{We will see that this assumption implies $F\sim L^2/G$, so that the conformal anomaly term is indeed smaller than the term we keep by a factor of $1/F \ll 1$.}

Ignoring the conformal anomaly, the $zz$ component of Einstein's equation gives
\be
 \left({\hat{a}_E' \over \hat{a}_E} \right)^2 - {\hat{a}_E^2 \over L^2} = - {8 \pi G \over z_0^4 } F {1 \over \hat{a}_E^2} \; ,
\ee
where $L$ is the AdS radius. So we have
\be
\int^{\hat{a}_E(z)}_{a_0} {da \over \sqrt{{a^4 \over L^2} - {8 \pi G F \over z_0^4}}} = z
\ee
where $a_0$ is the minimum value of $\hat{a}_E$ where $\hat{a}_E'=0$, 
\be
a_0 = {1 \over z_0} (8 \pi G F L^2)^{1 \over 4} \; .
\ee
For $z = z_0$, we have $\hat{a}_E \to \infty$, so we get
\be
\label{int1}
\int_{a_0}^\infty {da \over \sqrt{{a^4 \over L^2} - {8 \pi G F \over z_0^4}}} = z_0
\ee
which gives 
\be
\label{eqn}
F = {{\cal I}^4 L^2 \over 8 \pi G} \; .
\ee
where
\be
{\cal I} = \int_1^\infty {dx \over \sqrt{x^4-1}} = {\Gamma \left({3 \over 2} \right) \Gamma \left({1 \over 4} \right) \over  \Gamma \left({3 \over 4} \right)}  = {\sqrt{2} \over 2} K\left({\sqrt{2} \over 2}\right) \approx 1.311\; .
\ee
We recall that $F$ represents the energy density in units of the geometrical length scale in the conformal frame, where we have a CFT Minkowski space between two boundaries separated by $z_0$. As we discussed above, this would normally be expected to be of order the central charge $c_0$ of the CFT we are considering, but here we need it to take a much larger value $L^2/G \sim c_{3D}$, i.e. comparable to the central charge of the underlying holographic 3D CFT. However, such anomalously large values can arise with appropriate boundary physics, at least in certain holographic CFTs; see Appendix \ref{app:Casimir} for more discussion.

\subsubsection*{Lorentzian solutions with CFT matter}

It is straightforward to work out the Lorentzian cosmology that arises from this example. Again ignoring the conformal anomaly term, and using (\ref{eqn}), the Friedmann equation gives (in proper time coordinates)
\be
 \left({\dot{a} \over a} \right)^2 + {1 \over L^2} =   {{\cal I}^4 L^2 \over z_0^4 }  {1 \over a^4} \; .
\ee
This is equivalent to standard cosmological evolution with radiation and a negative cosmological constant. The universe expands from a big bang to a maximum scale factor and then contracts again. We can choose coordinates so that the maximum scale factor is $a=1$ when $\dot{a} = 0$. This gives $z_0 = {\cal I} L$. Defining $t = L s$ so that $s$ is the time in units of the curvature scale $L$, the evolution equation simplifies to
\be
 \left({\dot{a} \over a} \right)^2 + 1 =  {1 \over a^4} \; ,
\ee
with solution $a = \sqrt{\cos(2s)}$. In terms of the comoving time variable $t$, the metric is thus simply
\be
ds^2 = -dt^2 + \cos \left({2 t \over L} \right) d\vec{x}^2 \; .
\ee
so we have a big-bang / big-crunch cosmology with total proper age $L \pi/2$ determined by the AdS scale in the wormhole picture. 

As discussed in \cite{VanRaamsdonk:2021qgv}, the conformal anomaly terms can alter the behavior near the big bang and big crunch. For example, the special case where the conformal anomaly contributes only first-derivative terms to the Friedmann equation leads to an equation whose solution cannot be extended back past an initial time with energy density $T_{00} \sim 1/(c_0 G^2)$ where $c_0$ is a measure of the number of degrees of freedom in the effective field theory. 

\subsection{Non-conformal effective field theory}

In microscopic examples, we expect that the effective field theory describing quantum fields in our setup will be some non-conformal field theory.

Here, the stress-energy tensor computed on a fixed background will be some more complicated functional of the scale factor; in conformal coordinates we can represent this as
\be
T_{zz} = - {1 \over a^2} {3 \over z_0^4} F_a(z) \; .
\ee
The dependence of $F$ on $a$ may be complicated and non-local. In the cosmology picture, the resulting stress tensor evolution will also be more complicated and reflect the RG behavior of the field theory. In later work, we hope to investigate further the varieties of cosmological evolution associated with the vacuum stress tensors of more interesting field theories.

\subsection{Evolution with a scalar field}

It is natural to consider the possibility of one or more scalar fields with classical values that depend on time in the cosmology picture or on the $\tau$ coordinate of the wormhole. Such scalar field configurations preserve the spacetime symmetries of the setup and lead to models with time-dependent dark energy. In order to explain the current observations of an accelerating universe using our framework, having a time-dependent scalar field at some positive value of its potential is required, or at least is the simplest possibility. 

The presence of scalar fields is generic in the gravitational effective field theories that correspond to our microscopic models. According to the AdS/CFT correspondence, we have a light bulk scalar field for each low-dimension scalar operator in the CFT, with the scalar field mass related to the operator dimension (for the case of a 3D CFT) by\footnote{This gives a constraint $m^2 L^2 > -9/4$, which is the Breitenlohner-Freedman bound for 4D gravity theories with stable AdS vacua. It is not completely clear that such stability is required on our setup; it might be interesting to explore whether models based on more general gravitational theories (perhaps dual to unstable or non-unitary 3D CFTs) could be sensible.}
\be
\label{masses}
m^2 L^2 = \Delta (\Delta - 3) \; .
\ee
If the CFT is chosen so that the 4D cosmological constant $1/L^2$ is small, the scalar masses (which set the scale of the derivatives of the potential near the origin) will also be small; the result is that when these scalar fields vary, it is natural for them to be changing non-trivially throughout the spacetime.

If the CFT has both relevant and irrelevant scalar operators, the asymptotic values of the scalar fields will correspond to a saddle of the potential with both positive and negative directions. Generically, the potential will also have higher order terms, and these tend to make the potential positive for large values of the fields (assuming the AdS solution dual to our 3D CFT is stable). For example, in the 4D gauged supergravity models associated with 3D SCFTs with extended supersymmetry, the scalar potential is a quartic polynomial \cite{SUGRA4}. In models with less supersymmetry, the potential can be much more complicated (see e.g. \cite{Karndumri:2022rlf} and references therein for examples).

For solutions of the dual gravitational theory that are controlled by our 3D CFT in the UV, the most generic situation is to have scalar fields associated with the relevant or marginally relevant operators of the theory vary from the saddle-point values as we move away from the AdS boundary. These solutions correspond to taking the dual quantum field theory to be an RG flow obtained by perturbing the CFT with a relevant operator.\footnote{Even if we start with the strict 3D CFT, such perturbations might be induced by the coupling to other 3D CFT via the 4D CFT.} The strict CFT is a special case where all such perturbations are set to zero. 

In this section, we will describe the physical effects of such varying scalar fields in some generality, showing that they can lead to a phase of accelerated expansion. A more detailed analysis of these setups will be published elsewhere.

\subsubsection*{Scalar field evolution in the wormhole picture}

It is convenient to start with the radial evolution of the scalar field in the wormhole picture. Here, using proper distance coordinates, the equation of motion for scalar fields $\phi_i$ with potential $V(\vec{\phi})$\footnote{It will be convenient to absorb the cosmological constant into the definition of the scalar potential; we will assume this throughout this section.} is
\be
\label{wormdamped}
\phi_i'' + 3 H_E \phi_i' - \partial_i V(\vec{\phi}) = 0
\ee
(where $H_E \equiv a_E'/a_E$) and the field makes a contribution 
\be
\rho^\phi_E = -{1 \over 2} (\partial_\tau \vec{\phi})^2 + V(\vec{\phi})
\ee
to the right side of the scale factor evolution equation (\ref{FriedmannE}). In this picture, the evolution of $\vec{\phi}$ with $\tau$ is equivalent to the time evolution of a particle in a potential $-V(\vec{\phi})$ with damping coefficient $3 H_E$ that becomes constant in the asymptotically AdS regions. We are interested in time-symmetric solutions, so the derivative of $\vec{\phi}$ should vanish at the midpoint of the wormhole. Moving away from the time-symmetric point, $a_E$ increases,\footnote{The second time derivative of the scale factor is negative at the time-reversal symmetric point of the cosmology, so the second $\tau$ derivative is positive at the midpoint of the wormhole} so the evolution of $\vec{\phi}$ going away from this point will correspond to ordinary damped motion in the inverted potential.

For a given potential, the possible time-symmetric solutions will be in one-to-one correspondence with the value $\vec{\phi}_0$ of the scalar field at the midpoint of the wormhole. If $\vec{\phi}_0$ corresponds to an extremum of $V(\vec{\phi})$ the scalar will be constant throughout the geometry, and $V(\vec{\phi}_0)$ will be the cosmological constant. Otherwise, $\vec{\phi}$ will move downward (initially along the steepest path) in the inverted potential as we move toward the AdS boundary, as shown in Figure \ref{fig:scalar} (left).

\subsubsection*{Behavior at the AdS boundary}

As we have explained above, having a varying scalar field as we go to the asymptotically AdS regions corresponds to the underlying 3D CFT being deformed in the UV by a relevant or marginally relevant scalar operator.\footnote{We ignore the case where the scalar blows up at the asymptotically AdS boundary, since we are assuming that the UV behavior corresponds to some well-defined CFT.} In this case, the scalar will descend to a value $\vec{\phi}_B$ corresponding to an extremum of $-V$ (or rise to an extremum of $V$) at the boundaries where $V(\vec{\phi})$ is still negative.\footnote{In some cases, it might be that the scalar will oscillate before settling to this extremum. However, the relation (\ref{masses}) gives a lower bound $m^2 L^2 \ge -9/4$. At least in the asymptotic region where $a'/a = 1/L$, the equation (\ref{wormdamped}) for the scalar evolution corresponds to overdamped or critically damped motion, so there won't be oscillations in this region.} Alternatively, going away from the boundary, the scalar rolls down a direction in the potential $V(\vec{\phi})$ with negative second derivative, corresponding to a massless or negative mass-squared scalar that still satisfies the Breitenlohner-Freedman bound \cite{breitenlohner1982stability}.

\subsubsection*{Scalar field evolution in the cosmology picture}

In the cosmology picture, the scalars will take the values $\vec{\phi}_0$ at the time-symmetric point and have vanishing time derivatives here. 

The equation of motion in the cosmology picture with comoving (proper time) coordinates is
\be
\ddot{\phi}_i + 3H \dot{\phi}_i + \partial_i V(\vec{\phi}) = 0
\ee
and the field makes a contribution 
\be
\rho^\phi = {1 \over 2}  (\partial_t \vec{\phi})^2+ V(\vec{\phi})
\ee
to the energy density appearing on the right side of the Friedmann equation (\ref{Friedmann}). 

As is familiar from discussions of inflationary cosmology, the dynamics of the scalar field is that of a particle in a potential $V(\vec{\phi})$ with time-dependent damping coefficient $3H$, such that we have usual damping in the expanding phase of the universe and anti-damping in the contracting phase (or evolving backwards towards the big bang). 

In order to be stationary at the time-symmetric point, the scalars should be rolling up the potential before this. As we described above, a typical potential in a theory supporting a stable AdS vacuum will have interaction terms that make the potential bounded below and positive for large values of the field.

Thus, in the anti-damped motion going back toward the big bang, a typical situation would be for the scalar to descend to some lower value of the potential before rising again to positive values (time-reverse of Figure \ref{fig:scalar}, right).\footnote{The minimum value that the potential passes through might correspond to the endpoint of the RG flow discussed above, but with multiple scalars, the minimum value of the potential reached via the scalar field dynamics does not necessarily correspond to an extremum of the potential.} 

\subsubsection*{A phase of accelerated expansion}

Going backward from the time-symmetric point, we have a phase of decelerated expansion while the scalar potential is negative. But at earlier times when the scalar potential reaches positive values, we can have a phase of accelerated expansion. Whether or not such a phase exists depends on the details of the potential, and in particular whether the scale factor reaches $a=0$ going backwards in time before the scalar potential reaches large enough values to dominate over the matter and radiation. 

If the accelerated expansion in our universe is explained in this way, the scalar must be rolling sufficiently slowly at the $V > 0$ point corresponding to the present time to be consistent with the observations. In later work, we plan to investigate in detail whether there exist potentials giving rise to evolution that satisfies the observational constraints.




\subsubsection*{Inflation at early times?}

In our setup, the potential takes small values in the wormhole solution and for the late time cosmology (near the time-symmetric point) by construction (choosing an appropriate 3D CFT). However, in the surrounding ``landscape'' of string theory, the more generic values for the potential are expected to be closer to Planck scale. It is possible that the initial conditions for the matter/radiation dominated phase were set up by an early period of inflation where the scalar potential took much larger values, but it is not clear that such a phase is generic in our models or that it would happen at an energy scale where effective field theory is valid.

However, as we have emphasized above, it appears to be unnecessary in our model to understand the details of the early universe physics in order to compute cosmological perturbations. Cosmological observables in our model are computed most naturally in the wormhole picture where the effective field theory should be valid everywhere and the quantum state is time-independent. Here, the scalar field only explores a small region of the potential near the saddle that corresponds to the dual 3D CFT. 

So while there may be inflation in the early universe of the cosmology picture, and the physics of inflation may provide an explanation for the origin of perturbations, etc... in that picture, understanding the details of this early universe physics is not necessary to extract the predictions of the model. The wormhole picture provides in some sense a dual picture of the physics, with a different explanation for the origin of correlations relevant to cosmological observables. We will discuss this further in section 4 below.

\subsection*{Evolution from the initial time}

\begin{figure}
  \centering
  \includegraphics[scale=0.28]{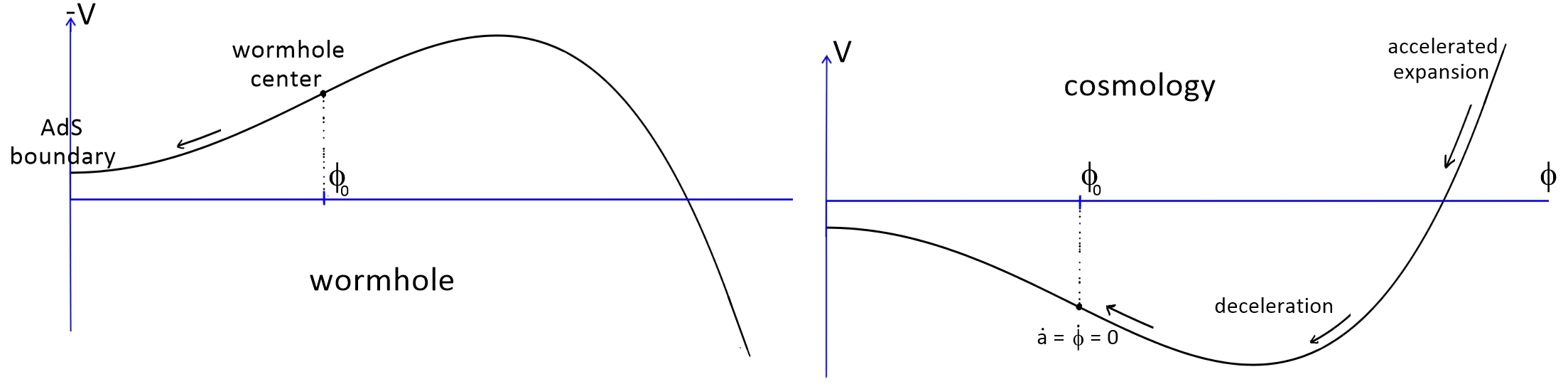}
  \caption{Left: evolution of the scalar from the wormhole center to the asymptotically AdS region corresponds to damped motion in the inverted potential $-V$ with damping ``constant'' $3 a_E'/a_E$. The evolution of the scalar is dual to the RG flow induced by perturbing the dual CFT by a relevant operator. Right: Evolution of the scalar field in the cosmology from early times to the time-reversal symmetric point corresponds to damped motion in the potential $V$ with damping constant $3 \dot{a}/a$. The initial positive values of the potential typically give rise to a phase of accelerated expansion before deceleration and collapse.}
\label{fig:scalar}
\end{figure}

In our scenario, it is most natural to consider the evolution of the scalar field from the asymptotically AdS boundary to the midpoint of the wormhole and then backward in time to the early universe as we have described. But it is also helpful to envision the evolution in the other direction, since this corresponds to more intuitive ordinary damped motion in the potential. 

The evolution is summarized in Figure \ref{fig:scalar}. Starting from a positive value of the potential and some large density of matter and radiation (that may have come from an inflationary phase) the scalar will roll down\footnote{A more general possibilitiy is that the scalar will move up the potential before rolling down, or have various oscillations.} to some negative value of the potential $V$ before rising again. The potential might oscillate up and down multiple times due to the motion of the scalar in the vicinity of a minimum. At the time-symmetric point, the scalar comes to rest at a point $\vec{\phi}_0$ between two extrema, approaching this point along an upward trajectory of steepest increase. 

The scalar field also takes the value $\vec{\phi}_0$ at the midpoint of the wormhole in the dual picture, but going away from this point evolves via damped motion in the inverted potential $-V$, eventually descending to the extremum that corresponds to our dual 3D CFT. 

The initial conditions required to get a time-symmetric scale factor (i.e. $\dot{\vec{\phi}} = 0$ when $\dot{a} = 0$) and end up in the wormhole picture at a non-minimal extremum of $-V$ would appear to be finely tuned in the cosmology picture, but the microscopic setup guarantees these will arise.

\subsubsection*{Symmetry breaking}

Having a non-zero scalar field expectation value away from the asymptotically AdS regions may be associated with a breaking of gauge symmetry and/or supersymmetry, such that the low-energy effective field theory relevant to the description of physics near the middle of the wormhole or in the cosmology picture is different than the effective field theory describing physics in the asymptotically AdS regions (which would be supersymmetric in microscopic models built from a 3D SCFT). If there is supersymmetry in the effective theory for the asymptotically AdS regions, a challenge is to explain why the scale of SUSY breaking is so much larger than the scale of the vacuum energy in the low energy effective description of the cosmology. This might may be explained fairly naturally if it is the value of the scalar field that sets the supersymmetry breaking scale, since the typical potentials in 4D gravity theories dual to CFTs remain of order $8 \pi G V \sim 1/ L^2$ when the scalar fields themselves have Planck scale variations.\footnote{For example, a typical Lagrangian density would be $(\partial \phi)^2 - c_0/(G L^2)  + c_2/L^2 \phi^2 +c_4  G/L^2 \phi^4$, with $c_i$ order one, so $8 \pi G T$ will stay of order $1/L^2$ as $\phi$ takes values of order $1/\sqrt{G}$.} As we have described above, if such scalar field expectation values and other features of the state are the only thing that breaks supersymmetry in the cosmology picture, the supersymmetric cancellations that prevent a large quantum correction to the cosmological constant may persist.

It would be useful to understand the possibilities for the low energy physics of theories obtained by starting from a supersymmetric 4D gravity with with an AdS vacuum and breaking symmetries via expectation values for scalars in directions corresponding to relevant operator deformations in the CFT.

\section{Fluctations about the background}

The Euclidean path integral that defines our model corresponds to a gravitational path integral; we have been assuming this will be dominated by a saddle-point configuration that corresponds to the background geometry we have been discussing. However, the path integral also includes fluctuations about the background geometry. In the cosmology picture, these fluctuations correspond to different possible realizations of structure in the universe. Thus, the microscopic wavefunction describes an ensemble of possible cosmologies, rather than a single instance.

We can ask about the size of the perturbations (e.g. via the variance of various quantities in the ensemble) and also about the correlations between perturbations in typical elements of the ensemble. Information about these cosmological perturbations is contained in the correlation functions of the stress-energy tensor and other fields, and ultimately in the Euclidean QFT path integral that defines the microscopic state.

For example, we can consider the two-point function of the difference of the energy density from background $\langle (\rho(x) - \bar{\rho})(\rho(0) - \bar{\rho}) \rangle$; the $x=0$ case gives the variance at a point. While the overall energy density expectation value $\bar{\rho}$ is spatially constant (since our setup preserves translation invariance), the individual configurations contributing to the wavefunction of the universe will be inhomogeneous, and the quantitative information about these inhomogeneities can be extracted from the two-point function.

As we have emphasized above, since the same microscopic Euclidean path integral that defines the state of the cosmology is associated with the vacuum state of the confining gauge theory, the fluctuations that give rise to cosmological structure have an alternative interpretation as vacuum fluctuations in the wormhole picture.  Correlations in the cosmology picture are related directly to the vacuum correlations in the wormhole picture. 

For example, in a realistic cosmology with a Maxwell field, correlation functions of the energy density of this field (which would tell us about CMB observables) are related by analytic continuation to correlations of the stress tensor component $T_{\tau \tau}$ for the same field in the wormhole picture. In particular, any energy density correlators at the time-symmetric point of the cosmology are simply equal to the vacuum correlators $T_{\tau \tau}$ at the midpoint of the wormhole, provided all the operators sit in the same two dimensional plane.\footnote{This is the plane formed by the two coordinates in $\mathbb{R}^3$ that are spatial in both pictures.}

In the wormhole picture, correlation functions of bulk fields should be related in a relatively simple way to the correlation functions of operators in the microscopic field theory, using an HKLL-type prescription appropriate to our background geometry. That is, the bulk operator should correspond to a linear combination of local operators in the underlying field theory (mostly localized to the pair of 3D CFTs). So one could in principle use the microscopic field theory to calculate cosmological observables, though in practice, employing the effective field theory description in the wormhole picture should be adequate for most questions.

In future work, we plan to investigate the behavior of the correlators in our model and understand the predictions for cosmology. However, there is already one  promising qualitative feature.

\subsubsection*{An alternative to inflation or a dual description?}

A distinctive feature in the density perturbations observed via the CMB is that there are correlations between regions of the universe that would have never been in causal contact if we assume simple cosmological evolution since the big bang. The theory of inflation naturally explains these correlations by introducing an early period of accelerated expansion. However, in our model, these correlations can be explained in a different way. They arise naturally since the cosmological correlators are directly related to vacuum correlators in the wormhole picture, and in a vacuum state on a static geometry, it is natural to have correlations at all scales (e.g. any two points have been in causal contact).\footnote{Here, it may be important that while the dual quantum field theory is confining, it still has massless Goldstone bosons associated with the breaking of global symmetry $G \times G \to G$. In a theory with electromagnetism, one of these light degrees of freedom corresponds to the integral $\int A$ of the electromagnetic field between the asymptotic boundaries. So the infrared description still has long-range correlations beyond the confinement scale.}

We note also that the standard horizon and flatness problems addressed by inflation are naturally explained by the $\mathbb{R}^3$ symmetry of the model and the fact that the natural "initial" state lives at the time-symmetric surface. On the other hand, we don't know whether perturbations in our model agree with cosmological observations at the quantitative level. 

While one possibility is that our model provides an alternative to inflation, it is also seems possible that the wormhole picture provides a dual description of the same physics. It may be that in the cosmology picture there is inflation and that this generates the density perturbations in the usual way. In this case, the results must agree with the calculations in the wormhole picture where no knowledge of the early-universe physics or the inflationary potential is required. In this scenario, the inflationary physics would be dual to the vacuum physics of the wormhole in a similar way to how field theory and gravity can provide different explanations for the same phenomenon in AdS/CFT.

\subsection{The wavefunction of the universe in the effective field theory description}

While the microscopic quantum state describing the cosmology is constructed using a CFT path integral, we can also think about the state of the universe from the effective field theory point of view as arising from a certain Euclidean gravity path integral in the effective field theory picture.

It is interesting to contrast this path integral with the one that Hartle and Hawking considered \cite{Hartle:1983ai}. Hartle and Hawking suggested that the wavefunction of the universe most naturally arises from a wavefunction $\Psi[g_0]$ where the probability amplitude for a field configuration $g_0$ is given by a Euclidean gravitational path integral over all four-dimensional Euclidean geometries / field configurations having $g_0$ as the data at a boundary, with no other boundaries allowed.

\begin{figure}
  \centering
  \includegraphics[scale=0.3]{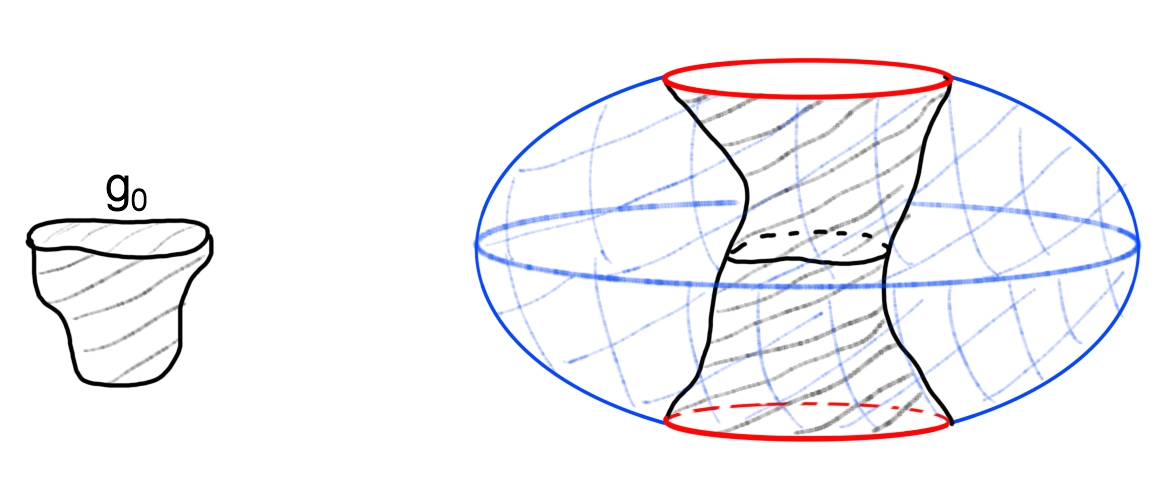}
  \caption{Left: A contribution to the Hartle-Hawking wavefunction of the universe. Right: A contribution to the gravitational path integral in our setup. In this case, the Euclidean gravity configurations have asymptotically AdS boundaries in the Euclidean past and future, and the fields at these boundaries are coupled together. The pictures correspond to the case of a closed universe; the flat universe we are focusing on can be obtained by taking the sphere volume to infinity.}
\label{fig:Wavefunction}
\end{figure}

In our approach, the state of the universe may also be naturally understood as arising from a Euclidean path integral. In the effective field theory description, the gravitational configurations contributing to the path integral used for computing observables have asymptotically AdS boundaries in the Euclidean past and future, and the fields in these asymptotic regions are coupled together via a non-gravitational CFT, as shown in Figure \ref{fig:Wavefunction}.
This non-gravitating field theory serves as an anchor for the fluctuting geometries contributing to the gravitational path integral. The holographic picture suggests that these geometries should be homologous to the fixed geometry of the non-gravitational field theory.

A qualitative difference between our case and the Hartle/Hawking picture is that the fields in our cosmological spacetime are in a mixed state, entangled with the fields in the non-gravitational quantum field theory. Tracing out these auxiliary fields may play a role in understanding how an ensemble of classical cosmologies arises from the wavefunction of the universe in our model.

\subsection{The arrow of time}
\label{sec:arrow}

The special state in our cosmology is time-symmetric, so one must understand how the arrow of time emerges for observers in the model. The same issue arises for the Hartle-Hawking proposal, and there have been a number of previous discussions (see e.g. \cite{Page:1985ei}). In \cite{VanRaamsdonk:2020tlr}, following Page \cite{Page:1985ei} we suggested that the model might describe an ensemble of different cosmologies, some of which have arrows of time in one direction and some of which have arrows of time in the other direction.  Here, we point out an interesting alternative, motivated by the idea that i) the arrow of time relates to the classical physics experienced by observers ii) the emergence of classicality involves decoherence and iii) that the physics of decoherence should occur either with forward quantum evolution or backward quantum evolution. The idea is that there might be two complementary bases of states, for which the basis elements are not time-reversal invariant, but the two bases are related by time-reversal. The in and out bases of states in discussions of scattering in quantum field theory provide an example. The full state can either be expressed in terms of the forward basis or the backward basis. Using the forward basis and tracing over some unobserved environmental degrees of freedom, we get an ensemble of states that might represent possible cosmologies with a forward arrow of time. Doing the same starting with the backward basis, we would then get an ensemble of possible cosmologies with a backward arrow of time. This picture has the intriguing property that each possible forward cosmology could be mathematically represented as a complicated superposition of backward cosmologies. However, further work is required to understand whether this possibility is sensible.

\section{Discussion}

In this paper, we have described a possible framework for cosmology motivated by two guiding principles (where the second was suggested by the first):
\begin{itemize}
    \item 
    There should be a fully non-perturbative microscopic description of the theory.
    \item
    There should be a preferred state.
\end{itemize}
In other words, we are exploring the most optimistic possible scenario for understanding the large-scale physics of our universe. 

The first principle, together with our current understanding of string theory, led us to focus on the possibility that the fundamental cosmological constant in the effective theory describing our universe is negative. The existence of an asymptotically AdS analytic continuation suggested both a fundamental description involving 3D CFTs and a preferred state constructed using a Euclidean path integral. An unexpected consequence is the existence of a dual physical system whose vacuum state encodes most of the cosmological physics.

The resulting universe has a number of attractive features. Homogeneity, isotropy, and flatness are consequences of the symmetry of the model (and the fact that the present state of the universe is most naturally understood via backward evolution from the time-symmetric point). The equivalence between cosmological observables and vacuum observables in the dual wormhole system provides a natural explanation for the existence of correlations at all scales. This relation also means that any cosmological observables can be computed in principle without knowledge of the physics near the big bang, and without detailed knowledge of the UV completion. In particular, even if there is inflation at early times in the cosmology description, we can compute cosmological perturbations in the dual picture without knowledge of the inflationary potential. 

The simplest toy example of our setup (with CFT matter) gives the simplest nontrivial $\Lambda < 0$ cosmology with radiation and cosmological constant leading to a big-bang / big crunch universe. This universe does not have an accelerating phase, but more generic examples of our setup have time-dependent scalar fields leading to time-dependent dark energy. These may exhibit a phase of accelerated expansion before the collapse and a phase of matter/radiation domination before that.

The small cosmological constant is related to the large number of degrees of freedom in the underlying 3D CFTs and (likely) the presence of supersymmetry in the fundamental description. The gauge group for the effective theory is controlled by the global symmetry group of the underlying 3D CFT; microscopic supersymmetric examples with almost any global symmetry group are known. Finally, both the model and the state are completely specified by our choice of 3D and 4D CFTs.

A natural next step (that we are currently pursuing) is to investigate in detail the cosmological evolution and cosmological correlation functions predicted by specific examples of this type of model, using an effective field theory approach, in order to understand whether realistic examples can be obtained.

\subsection{Challenges}

We emphasize that it is still an open question to demonstrate definitively that there are fully microscopic examples of this cosmological scenario, and a further challenge to understand whether examples of such models make predictions that are consistent with observations at a quantitative level. 
An important question is to verify that 3D-4D-3D quantum field theory systems of the type we have described can be dual to a Lorentzian wormhole. To demonstrate the viability of these models more directly, there are various options. For the fully microscopic examples suggested in \cite{VanRaamsdonk:2021qgv}, a classical type IIB supergravity calculation (or alternatively, a strongly coupled field theory calculation) is required to verify that the dual geometries have the right properties. Alternatively, one could try to understand precisely the four-dimensional effective gravitational theory for a microscopic example and understand how this can support a wormhole. Better understanding when and how quantum field theories can exhibit the appropriate vacuum energies to support a static Lorentzian planar wormhole is clearly an important step.

To understand whether our scenario can give rise to realistic examples of cosmology, there are many things to check. These include: whether the accelerated expansion from scalar potentials of the type we have described can satisfy the observational constraints on the time variation of dark energy,\footnote{As we have explained, for the effective gravitational theories arising from holography, it is natural for the derivatives of the scalar potential to be small when the cosmological constant is small, so this is somewhat plausible.} whether the cosmological perturbations in our models can have the right magnitude and correlations, and whether we can satisfy numerous other phenomenological constraints, e.g. obtaining Standard Model as an effective theory, getting the right amount of dark matter and baryon asymmetry,\footnote{The time-dependent scalar fields that are generic in this class of models may be helpful in explaining the baryon asymmetry \cite{Affleck:1984fy}.} etc... . Since accelerated expansion in our models relies on having one or more time-dependent scalar fields with a very flat potential, it must be understood why other effects of such light scalar fields (e.g. long range forces or time-dependent couplings) have not been observed.\footnote{A simple possibility is that these scalar fields are in the dark sector; fermions coupling to the scalar via $\phi \bar{\psi} \psi$ couplings in the underlying theory will have large masses when $\phi$ has large expectation values. For a more detailed discussion, see e.g. \cite{carroll1998quintessence,amendola2000coupled}.}

Regardless of whether the scenario is realistic, it may still allow us to make progress on fundamental questions about quantum cosmology, just as we have learned about the quantum description of black holes through AdS/CFT.

\subsection{Cosmological Islands}

The fact that the effective field theory description of our cosmology includes an auxiliary system with which the fields in the cosmological spacetime are entangled is related in an intriguing way to observations in \cite{Hartman:2020khs}. There, the authors asked when regions of cosmological spacetimes could be ``entanglement islands'', regions whose degrees of freedom are entangled with some other disconnected system in the effective field theory such that the two parts together form a proper quantum subsystem with a generalized entropy (and thus potentially dual to a quantum subsystem of a microscopic theory). They found that for certain $\Lambda < 0$ flat cosmologies, large enough spatial systems at times close enough to the recollapse point satisfy the criteria to be an island. In that analysis it was not clear what the complementary disconnected system or the microscopic description could be. Our analysis suggests that exactly in the case where \cite{Hartman:2020khs} found islands, we can give a microscopic description, and the microscopic description indeed involves an auxiliary system with which the fields in the cosmology picture are entangled.

We can also argue directly that large enough subsystems of our cosmology are islands. Consider a ball-shaped region of radius $R$ in the microscopic 4D CFT. For large enough $R$, this will encode not only the physics of the same region in the 4D CFT in the effective description, but also part of the cosmology. To see this, consider a ball of radius $R$ in the auxiliary system on the time-symmetric slice. This will have thermal entropy scaling like $R^3$. When $R$ is much larger than the thermal scale, the fields in this region will be mostly purified by fields in the cosmology picture. It should thus be possible choose a similar ball-shaped region of the cosmology (also on the time-symmetric slice) with size proportional to $R$ such that the entropy of the combined system now scales as $R^2$, coming from degrees of freedom near the boundaries of the two balls. Thus, for large enough $R$, this combined system with a cosmological island has smaller generalized entropy and provides the correct dual description of the microscopic CFT region. 

This is even clearer when the 4D CFT is holographic. In this case, an RT surface for a ball-shaped region in the microscopic theory that remains outside the horizon will have area scaling like $R^3$ for large $R$, while an RT surface that enters the horizon and ends on a spherical region of the ETW brane will have area that scales like $R^2$ for large $R$.\footnote{This $R^2$ may come with a large factor, since the ETW branes with localized gravity are associated with regions of the internal space with large volume in a higher-dimensional picture. So in the RT formula in the effective five dimensional theory, there will be terms proportional to the three-dimensional area of the intersection with the ETW brane, with a large coefficient.} So for large enough $R$, the latter surface is preferred, and the entanglement wedge includes part of the cosmology. This transition in entangling surfaces and the appearance of islands on the cosmological brane was originally observed and studied in \cite{Cooper2018}.

We can also take the limit where the 4D CFT region size goes to infinity, in which case the entire time-symmetric slice in the cosmology is an island. Hence, the 4D CFT state at the time-symmetric point already has enough data to encode the entire cosmological evolution without any CFT time evolution. This is consistent with the cosmological time evolution being emergent and unrelated to the 4D CFT time evolution.\footnote{The disconnection between cosmological time evolution and CFT time evolution is even more evident if we consider that islands at the time-symmetric point $t=0$ of the cosmology can be part of the entanglement wedge of regions $R$ of the 4D auxiliary CFT even at late CFT times $t_{CFT}>0$ \cite{Hartman:2020khs}.} Moreover, in the special case where the 4D CFT is holographic, the Wheeler-DeWitt patch of the time-symmetric point includes the entire end-of-the-world brane trajectory, in agreement with the general analysis.

It is interesting to ask which other cosmological spacetimes can support islands - this may give us a hint whether any other types of cosmologies might have a holographic description similar to the one we are considering. This question was considered in some generality in \cite{Hartman:2020khs} and \cite{Fallows:2021sge,bousso2022islands}. For $\Lambda < 0$ ball-shaped islands continue to be possible for closed or open universes (i.e. where the spatial curvature is positive or negative), provided that the radius of curvature is sufficiently large \cite{bousso2022islands}. This is consistent with the fact that the microscopic approach suggested in this paper can be generalized to positive or negative spatial curvature by replacing the $\mathbb{R}^3$ in the construction with $S^3$ or $H^3$. The curvature scale for these must be sufficiently large compared to the size of the interval separating the 3D CFTs to end up with the desired confining phase \cite{Cooper2018}. 

For $\Lambda \geq 0$, it was found that no islands are possible in flat \cite{Hartman:2020khs} or open \cite{bousso2022islands} cosmologies. For closed cosmologies (i.e. with positive spatial curvature), a new island candidate is given by the whole cosmological universe, regardless of the sign of $\Lambda$. Thus, it is interesting to ask whether some special state of a 4D CFT on $S^3$ might also encode the physics of cosmological spacetimes with $S^3$ spatial geometry and  $\Lambda \ge 0$. It is not clear how such a state could arise from a Euclidean path integral, though. 

\subsection{Summary of features}

Whether or not the models we are describing turn out to be viable for realistic cosmology, they have a number of interesting features that might be useful in future approaches. To conclude the paper, we highlight some of the particular features that appear and that might inform future research.

First, we have various motivations to consider $\Lambda < 0$ cosmology:
\begin{itemize}
\item
We take the existence of a complete, non-perturbative microscopic description as a guiding principle for describing cosmology. Given the present state of our understanding of string theory, this suggest that we focus more on the $\Lambda < 0$ possibility. 
\item
Specifically, with $\Lambda < 0$, we avoid the vacuum instabilities and complicated bubbling multiverse picture that appears in $\Lambda > 0$ discussions. While this picture may dynamically give rise to anthropically allowed positive values of $\Lambda$ if enough $\Lambda > 0$ string vacua exist \cite{Bousso:2000xa}, it seems difficult to establish that the required landscape of vacua exists (see e.g. \cite{Danielsson:2018ztv}), and difficult to imagine what a microscopic description of such a scenario could look like. 
\item
Considering models with time-dependent scalar fields (which can lead to an accelerated expansion phase even if $\Lambda<0$) is natural in that this is the most general situation consistent with the symmetries of cosmological spacetimes. It is also natural from the string theory point of view, since the best understood regions of the string landscape with small positive potential energy are regions near AdS minima where we have excited some scalar fields. 
\end{itemize}
An interesting feature of $\Lambda < 0$ is that it suggests there may be a natural state for the cosmology constructed from a Euclidean theory:
\begin{itemize}
\item
Many natural $\Lambda < 0$ cosmological backgrounds have time-reversal symmetry about the $\dot{a} = 0$ point, and these time-symmetric geometries have an analytic continuation to a reflection-symmetric Euclidean geometry. The corresponding Euclidean gravity theory can be used to define a natural state for the cosmology. For this choice:
\item 
The time-reversal symmetry of the background spacetime extends to the quantum state.
\item
The cosmological observables (for the full quantum state describing the ensemble of possible cosmologies) are analytic functions of coordinates.
\item
The late time cosmological observables can be calculated precisely without a detailed knowledge of the big bang or initial conditions in the cosmology picture.
\item
The cosmological observables are related to the vacuum observables in a dual picture. This naturally implies correlations between regions of the universe that were never in causal contact.
\end{itemize}
There are interesting features related to the holographic description:
\begin{itemize}
\item
Three-dimensional holographic quantum field theories play an essential role in the description of the physics. Since these provide descriptions of vacuum AdS solutions, it is natural that they would continue to play an important role if we add a non-zero density of matter/radiation/scalar potential, etc... . In our construction, the three-dimensional theory no longer appears in the physical degrees of freedom of the holographic description, but still appears in the path integral that constructs the state. 
\item
The fundamental microscopic degrees of freedom housing the quantum state of the universe are not conventionally holographic. In our example, they are a small $N$ CFT. These holographic degrees of freedom are not ``at the boundary'' of the cosmological spacetime. Using the language of recent discussions of black hole evaporation, the cosmological spacetime is an ``island.'' Related to this, we have that:
\item
The particles/fields in the effective description of cosmology are entangled with an auxiliary system that does not interact directly with the cosmology. Tracing out these auxiliary degrees of freedom, we have a mixed state of the cosmological effective field theory, describing an ensemble of different cosmologies.
\item
The full history of the universe is encoded in the state of the underlying 4D CFT at a single time, and should not be affected by changes to the CFT Hamiltonian at earlier or later times. Thus, time evolution in the cosmology seems unrelated to time evolution in the underlying CFT. We can say that time in the cosmology is emergent. 
\item
Finally, the scalar field potential in the effective theory is directly related to properties of the underlying 3D CFT. Taking the number of CFT degrees of freedom to be very large (as would seem appropriate for describing the entire universe) naturally leads to both a small cosmological constant and small derivatives for the scalar potential.
\end{itemize}
In special cases where the 4D field theory may also be holographic, we have that:
\begin{itemize}
\item
The cosmological spacetime lives behind the horizon of a black hole, and the cosmological singularities are related to the black hole initial and final singularities. 
\item
In our specific setup, the cosmological physics lives on a brane in a higher-dimensional spacetime. 
\item
Gravity is localized to the brane because the 3D CFTs associated with the four-dimensional physics have many more local degrees of freedom than the 4D CFT associated with the five-dimensional phsyics.
\end{itemize}
Finally, the models have some interesting consequences for phenomenological issues:
\begin{itemize}
\item
The only input to the model is the choice of the underlying coupled 3D/4D field theory. The gauge group of the effective theory for the cosmology is determined by the global symmetry of the 3D CFT. 
\item
The quantum state in the cosmology, and thus the cosmological perturbations are fixed given the choice of Euclidean effective field theory.
\item
Correlations between regions of the universe that were never in causal contact arise naturally because they are related to vacuum correlators in a dual picture.
\item
If the microscopic 4D CFT (and the dual gravity theory) is supersymmetric, this supersymmetry is only broken by the state in the cosmology picture. This may allow supersymmetric cancellations in the leading contribution to the cosmological constant to persist and explain why the cosmological constant can remain small. In the wormhole picture, supersymmetry is broken explicitly, but only by nonlocal effects (the incompatibility of the two boundary conditions). Again, this seems to avoid large corrections to the cosmological constant.
\item
The Friedmann equation implies that the total energy density at the time-symmetric point vanishes, so the magnitude of the (negative) dark energy is equal to the magnitude of the remaining energy. The time scale for the variation of these quantities is the age of the universe, so the dark energy and other sources of energy are expected to have a similar order of magnitude for most of the age of the universe. This may help explain the cosmological coincidence problem.
\end{itemize}

\section*{Acknowledgements}

We would like to thank Panos Betzios, Matt Kleban, Lampros Lamprou, Henry Maxfield, Yasunori Nomura, Liam McAllister, Douglas Scott, Eva Silverstein, Chris Waddell, David Wakeham, and Aron Wall for helpful discussions and comments. This work is supported in part by the National Science and Engineering Research Council of Canada (NSERC) and in part by the Simons foundation via a Simons Investigator Award and the ``It From Qubit'' collaboration grant. PS is supported by an NSERC C-GSD award. This work is partially supported by the U.S. Department of Energy, Office of Science, Office of Advanced Scientific Computing Research, Accelerated Research for Quantum Computing program ``FAR-QC'' (SA) and by the AFOSR under grant number FA9550-19-1-0360 (BS).

\appendix

\section{States from sliced Euclidean path integrals}
\label{app:slicing}

In this appendix, we review the construction of Lorentzian states from slicing Euclidean path integrals.

Consider a Euclidean quantum field theory on a space $M$ such that the background and the theory are refection-symmetric about some surface $A$ which divides $M$ into two equivalent parts $M_\pm$. The partition function of this Euclidean theory can be defined by the path integral
\be
Z = \int_M [d \phi] e^{-S_E} \; ,
\ee
where $S_E$ is the Euclidean action for the theory. We can use this path integral to define a particular state of the field theory degrees of freedom on the surface $A$. The wavefunctional for this state is
\be
\Psi_A[\phi_0] = {1 \over Z^{1 \over 2}} \int_{M_-}^{\phi(A) = \phi_0} [d \phi] e^{-S_E} \; ,
\ee
so the probability amplitude for a particular field configuration $\phi_0$ is a sum over all field configurations on $M_-$ with the fields on $A$ set to $\phi_0$. Observables in this state at time 0 are computed by inserting operators on the surface $A$ in the original path integral. More generally, certain time-dependent observables for this state in the associated Lorentzian quantum field theory are related by analytic continuation to observables in the Euclidean quantum field theory.

In the special case where $M$ has a non-compact direction with a translation symmetry and $A$ is a surface that divides the space in half perpendicular to this direction, the state $\Psi_A$ is the vacuum state of the associated Lorentzian quantum field theory (which has a time-independent Hamiltonian in this case). The reason is that the vacuum state can be defined, up to normalization, as $\lim_{\beta \to \infty} e^{- \beta H} |\Psi_0 \rangle$ (where $|\Psi_0 \rangle$ is any state that has overlap with the vacuum) and the path integral version of this expression is the sliced Euclidean path integral we have described.

\section{Cosmological equations in conformal coordinates}
\label{app:conformal}

For some applications, it is useful to consider the evolution equations that determine the background cosmology and wormhole spacetimes in conformal coordinates in which the metric is a scaled version of Minkowski space.

\subsubsection*{Background geometry and stress-energy tensor: conformal time}

In the cosmology picture, the background metric with conformal time is
\be
\label{FRWconf}
ds^2 = \hat{a}^2(\eta)( -d \eta^2 + dw^2 + dx_i dx_i) \; .
\ee
Here $\hat{a}(\eta) = a(t)$ and $t$ and $\eta$ are related by $d\eta / dt = a^{-1}(t)$. In these coordinates, the stress-energy tensor is
\be
T^{\eta \eta} = {1 \over \hat{a}^2(\eta)} \hat{\rho}(\eta) \qquad T^{ij} = {1 \over \hat{a}^2(\eta)} \delta^{ij} \hat{p}(\eta) \qquad T^{ww} = {1 \over \hat{a}^2(\eta)} \hat{p}(\eta)
\ee
where $\hat{\rho}(\eta) = \rho(t)$ and $\hat{p}(\eta) = p(t)$.

In these coordinates the Friedmann equation gives
\be
{1 \over \hat{a}^4} \left({d \hat{a} \over d \eta}\right)^2 + {1 \over L^2} = {8 \pi G \over 3} \hat{\rho}
\ee
and the conservation equation is
\be
\label{Econs2}
\dot{\hat{\rho}} = -{3 \over \hat{a}} {d \hat{a} \over d \eta} (\hat{\rho} + \hat{p}) \;
\ee

\subsubsection*{Wormhole picture: conformal distance coordinates}

It will also be convenient to describe the wormhole in terms of a conformal distance coordinate $z$, in terms of which the metric is
\be
\label{worm2}
ds^2 = \hat{a}_E^2(z)( dz^2 - d \zeta^2 + dx_i dx_i) \; .
\ee
The scale factor $\hat{a}_E(z)$ is related to $\hat{a}(\eta)$ by analytic continuation $\eta^2 \leftrightarrow -z^2$.

In these coordinates, the range of $z$ is finite, $z \in [-z_0,z_0]$. The coordinate locations $\pm z_0$ correspond to the asymptotically AdS boundaries; the scale factor $\hat{a}_E(z)$ has simple poles at these locations.

The stress-energy tensor in this picture is
\be
T^{zz} = - {1 \over \hat{a}_E^2(z)} \hat{\rho}_E(z) \qquad T^{\zeta \zeta} = - {1 \over \hat{a}_E^2(z)} \hat{p}_E(z) \qquad T^{ij} = {1 \over \hat{a}_E^2(z)} \delta^{ij} \hat{p}_E(z)
\ee
where $\hat{\rho}_E$ and $\hat{p}_E$ are related to $\rho$ and $p$ by analytic continuation $\eta^2 \leftrightarrow -z^2$.

The analog of the Friedmann equation  in this picture is
\be
-{1 \over \hat{a}_E^4} \left({d \hat{a}_E \over d z}\right)^2 + {1 \over L^2} = {8 \pi G \over 3} \hat{\rho}_E
\ee
and the conservation equation is
\be
\label{Econs4}
{d \hat{\rho}_E \over d z} = -{3 \over \hat{a}_E} {d \hat{a}_E \over d z} (\hat{\rho}_E + \hat{p}_E) .
\ee

\section{Requirement of negative integrated null energy in the wormhole picture}

The requirement of negative null energy at the midpoint of the wormhole can be extended to a requirement for integrated null energy between the two asymptotically AdS boundaries \cite{Galloway:1999bp}.

Considering the expression
\be
\label{defN}
N \equiv 8 \pi G \int d \lambda T_{\mu \nu} {dx^\mu \over d \lambda} {dx^\nu \over d \lambda} \; .
\ee
and taking $x(\lambda)$ to be a null geodesic between the two asymptotically AdS boundaries with $\lambda$ the affine parameter, we find
\be
N = 8 \pi G \int {dz \over \hat{a}_E^2(z)} T_{++}(z) = 8 \pi G \int dz T^{--}(z) \;.
\ee
Using Einstein's equations, we have that
\be
8 \pi G T_{++} = R_{++} = 4 \left( {\hat{a}_E' \over \hat{a}_E} \right)^2 - 2 {\hat{a}_E'' \over \hat{a}_E} \; .
\ee
Then we can check that
\bea
N &=& 2 \int dz \left\{ {d \over dz} \left[{1 \over \hat{a}_E}{d \over dz}{1 \over a} \right] - \left[{d \over dz}{1 \over \hat{a}_E}\right]^2 \right \} \; \cr
&=&   -  \int_{-z_0}^{z_0} dz \left[{d \over dz}{1 \over \hat{a}_E}\right]^2 < 0 \; ,
\eea
where $\pm z_0$ are the locations of the two asymptotically AdS boundaries. Near these points, $1/\hat{a}_E(z)$ behaves as $C(z_0 \mp z)$, so the total derivative term in the first line term vanishes.

We conclude that the null energy integrated from one side of the wormhole to the other as in (\ref{defN}) must be negative. This is not possible for any ordinary matter satisfying the null energy condition, but it is possible for the vacuum energy of quantum fields, and this is the situation we are considering.

\section{Negative Casimir energies in quantum field theory}
\label{app:Casimir}

In this appendix, we briefly recall a few relevant aspects of negative Casimir energies in quantum field theory. 

\subsubsection*{Negative energies with a periodic direction}

Consider any quantum field theory on $\mathbb{R}^{3,1}$ that is conformal in the UV (including UV-free theories). If we consider this quantum field theory at finite temperature, the energy density relative to the vacuum state at temperatures much higher than any other scale will behave as 
\be
T_{00} = c_T T^4
\ee
where $c_T$ gives some measure of the number of UV degrees of freedom. The full stress-energy tensor in this high-temperature limit will be approximately traceless (again because of the UV CFT), so we will have
\be
T_{ij} = {1 \over 3} \delta_{ij} c_T T^4 \; .
\ee
The thermal state of this CFT may be constructed by considering the Euclidean version of the quantum field theory with a Euclidean time direction that is periodic, with anti-periodic boundary conditions for fermions. The thermal density matrix for the original theory can be obtained by slicing the Euclidian path integral at some Euclidean time along the thermal circle.\footnote{Alternatively, we can slice the path integral into two pieces at a pair of times on opposite sides of the circle to define the thermofield double state.}

We can slice this same path integral perpendicular to one of the $\mathbb{R}^3$ directions to define the vacuum state of the field theory on $\mathbb{R}^2 \times S^1$ with antiperiodic boundary conditions for fermions on the $S^1$. 

The Casimir stress-energy tensor for this vacuum state (i.e. the stress-energy relative to its value on Minkowski space) is directly related to the thermal stress-energy tensor above, since both are obtained by inserting an operator into the Euclidean path integral that defines these two states. 

Specifically, if $\tau$ is the periodic coordinate and $\eta$ is the time coordinate in the theory with a periodic direction, we have 
\beas
T^{Cas}_{\tau \tau} &=& - c_T {1 \over \beta^4} \cr
T^{Cas}_{\eta \eta} &=& -{1 \over 3} c_T {1 \over \beta^4} \cr
T^{Cas}_{ij} &=& \delta_{ij} {1 \over 3} c_T {1 \over \beta^4}
\eeas
Thus, we have a negative energy density $T^{Cas}_{\eta \eta}$ and a negative null energy in the periodic direction of $T_{++} = T_{--} = -4/3 c_T \beta^{-4}$. 

These energy densities can be made arbitrarily negative by taking the length $\beta$ of the periodic direction to be small.

\subsubsection*{Enhanced negative energies in systems with boundaries}

In order to support the wormhole geometries that could give rise to realistic cosmology, we need negative energy that extends over a spatial region that is much larger than the scale of the negative energy itself. A naive way to construct this would be to take this quantum field theory state from the example above, consider the reduced state on a subsystem $\tau \in (0, \beta - \epsilon)$ of the periodic direction, and try to purify this in a system where the $\tau$ direction has a larger extent. 

It is impossible to purify this system to a translation-invariant state of a quantum field theory on Minkowski space, since the averaged null energy condition (ANEC) \cite{Faulkner:2016mzt,Hartman:2016lgu} says that integral of $T_{++}$ over a complete null geodesic must be non-negative. So in any purification on Minkowski, we must have regions of positive null-energy surrounding this region of negative null energy. However, we are not aware of any restriction on the width of a region where $T_{++}$ is bounded above by a certain negative value. The holographic models of \cite{VanRaamsdonk:2021qgv,May:2021xhz}, which have interfaces that create a finite spatial extent for the CFT, provide examples to suggest that there is no such restriction. It would be interesting to understand whether even weakly coupled quantum field theories can exist in states with a negative energy much larger than the geometrical scale (distance between interfaces/boundaries) for the right choice of boundary physics.

\bibliographystyle{jhep}
\bibliography{references}

\end{document}